\documentclass[journal,12pt,onecolumn,draftclsnofoot,twoside]{IEEEtran}
\normalsize

\usepackage[utf8]{inputenc}
\usepackage[noadjust]{cite}

\usepackage{graphicx}
\usepackage{pstricks}
\usepackage{pgfplots}
\usepackage[caption=false,font=footnotesize]{subfig}

\usepackage{amsmath}
\usepackage{amssymb}
\usepackage{mathtools}
\usepackage{xcolor}
\usetikzlibrary{shapes,arrows,positioning,calc,patterns}

\usepackage{enumitem}
\usepackage{bm}

\newtheorem{theorem}{Theorem}
\newtheorem{lemma}{Lemma}
\newtheorem{proposition}{Proposition}
\newtheorem{remark}{Remark}

\allowdisplaybreaks

\DeclareMathOperator{\snr}{snr}
\DeclareMathOperator{\mmse}{mmse}
\DeclareMathOperator{\E}{\mathbb{E}}
\DeclareMathOperator{\var}{Var}
\DeclareMathOperator{\sign}{sign}
\DeclareMathOperator{\dr}{d\hspace{-0.05cm}}

\usepackage[normalem]{ulem}

\begin{document}

\setlist[description]{font=\normalfont\itshape\space}

\title{Nonparametric Estimation of the Fisher Information and Its Applications}

\author{Wei~Cao,~\IEEEmembership{Student~Member,~IEEE},
       Alex~Dytso,~\IEEEmembership{Member,~IEEE},
       Michael~Fau{\ss},~\IEEEmembership{Member,~IEEE},
       H.~Vincent~Poor,~\IEEEmembership{Fellow,~IEEE},
       and~Gang~Feng,~\IEEEmembership{Senior~Member,~IEEE}
\thanks{This work was supported in part by the U.S. National Science Foundation under Grants CCF-0939370, CCF-1513915, and CCF-1908308, and in part by the German Research Foundation (DFG) under Grant 424522268. 
This paper was presented in part in \cite{cao2020nonparametirc}.
}
\thanks{W. Cao and G. Feng are with the National Key Lab of Science and Technology on Communications, University of Electronic Science and Technology of China, Chengdu 611731, China 
(email: wcao@std.uestc.edu.cn, fenggang@uestc.edu.cn).}
\thanks{A. Dytso, M. Fau{\ss}, and H. V. Poor are with the Department of Electrical Engineering, Princeton University, Princeton, NJ 08544, USA (email: adytso, mfauss, poor@princeton.edu).}%
}

\maketitle

\vspace{-0.2in}

\begin{abstract}
This paper considers the problem of estimation of the Fisher information for location from a random sample of size $n$.  First, an estimator proposed by Bhattacharya is revisited and improved  convergence rates are derived. Second, a new estimator, termed a clipped estimator, is proposed.  Superior upper bounds on the rates of convergence can be shown for the new estimator compared to the Bhattacharya estimator, albeit with different regularity conditions. Third, both of the estimators are evaluated for the practically relevant case of a random variable contaminated by Gaussian noise. Moreover, using Brown's identity, which relates the Fisher information and the minimum mean squared error (MMSE) in Gaussian noise, two corresponding consistent estimators for the MMSE are proposed. Simulation examples for the Bhattacharya estimator and the clipped estimator as well as the MMSE estimators are presented. The examples demonstrate that the clipped estimator can significantly reduce the required sample size to guarantee a specific confidence interval compared to the Bhattacharya estimator.
\end{abstract}


\IEEEpeerreviewmaketitle

\newpage
\section{Introduction}

\IEEEPARstart{T}{his} work considers the problem of estimating the \emph{Fisher information} for location of a probability density function (PDF) $f$ based on $n$ random samples $Y_1, \ldots, Y_n$ independently drawn from $f$. To clarify, the Fisher information of $f$ is given by 
\begin{align}
    I(f) = \int_{t\in\mathbb{R}} \frac{(f'(t))^2}{f(t)} \dr t, \label{eq:Fisher:Info:Def}
\end{align}
where $f'$ is the derivative of $f$. 

Estimation of the Fisher information in \eqref{eq:Fisher:Info:Def} was first considered by Bhattacharya  in \cite{bhattacharya1967estimation}, where a large sample regime was studied. In \cite{bhattacharya1967estimation}, a \emph{plug-in estimator} was proposed based on estimates of $f$ and $f'$ obtained via the \emph{kernel method}. Amongst other things, the work in \cite{bhattacharya1967estimation} produced error bounds on the estimation of the density and its derivative, and, under some regularity conditions, the proposed Fisher information estimator was shown to be consistent. 

Estimation of the derivatives of PDFs is important for plug-in methods.  In particular, kernel based methods for estimation of the derivatives of a PDF have received considerable attention. For example, the work of Schuster \cite{schuster1969estimation} considered estimation of higher-order derivatives of a PDF and has shown that, under mild regularity conditions, the estimation error for the higher-order derivatives can be controlled by the estimation error for the corresponding \emph{cumulative distribution function} (CDF). The interested reader is referred to \cite{ruschendorf1977consistency,silverman1978weak,roussas1991kernel,wertz1979statistical,tsybakov2004introduction} and references therein.   

As previously mentioned, the estimation of the Fisher information was first considered in \cite{bhattacharya1967estimation}. The bounds of \cite{bhattacharya1967estimation} have been revised by Dmitriev and Tarasenko in \cite{dmitriev1974estimation}. The work of \cite{dmitriev1974estimation} was also the first to consider the problem of entropy estimation. The techniques of \cite{bhattacharya1967estimation} and \cite{dmitriev1974estimation} have been generalized by Nadaraya and Sokhadze in \cite{nadaraya2016integral} to functionals that depend on the first $m$th derivatives of the density. In this work, we will recover the rates of [10] with less restrictive assumptions.

In \cite{donoho1988one}, Donoho has shown that in general, without making any assumptions on the density, the estimation of the Fisher information is a one-sided inference problem. More precisely, the true Fisher information cannot be upper  bounded based on samples alone and can only be lower bounded by minimizing Fisher information over a suitably chosen set of densities.
Moreover, \cite{donoho1988one} has also proposed a two-step procedure for estimating the Fisher information. In the first step, the empirical CDF is computed. In the second step, the smallest Fisher information attained on the ball centered at the empirical CDF, where the radius of the ball is defined via the Kolmogorov distance, is computed. Finally, the computed Fisher information is used as an estimate of the actual Fisher information. This method is closely related to the method of Huber splines \cite{huber1974fisher}. 

Estimation of the \emph{parametric Fisher information}\footnote{ Let $\{ f(x; \theta) \}, \theta\in \Theta$ denote an indexed set of PDFs, the parametric Fisher information is given by $I(\theta)=\int_{t \in \mathbb{R}} \frac{ (\partial_{\theta} f(t;\theta) )^2}{  f(t;\theta)} {\rm d} t $. The definition of the Fisher information in \eqref{eq:Fisher:Info:Def} agrees with the parametric one for the shift family, \textit{i.e.}, $f(x; \theta) = f(x- \theta)$.} has also received some attention in the literature. Particularly, Spall in \cite{spall2005monte} proposed to use a plug-in method by first performing nonparametric density estimation by perturbing each of the experiments followed by numerical gradient computation and followed by averaging. A non plug-in method was shown by Berisha and Hero in \cite{berisha2014empirical} where it was proposed to estimate an $f$-divergence and then estimate the parametric Fisher information by using the fact that $f$-divergences locally behave like the parametric Fisher information. We note, however, that the estimation of the parametric Fisher information and the Fisher information in \eqref{eq:Fisher:Info:Def} are typically different in spirit and purpose. On the one had, estimation of the parametric Fisher information typically assumes that the pdf of the model is known and the estimation procedure is typically performed as an alternative to the integration (i.e., Monte Carlo simulation). On the other hand,  estimation of the Fisher information in \eqref{eq:Fisher:Info:Def} does not assume the knowledge of the pdf and the goal is to estimate the Fisher information of unknown distribution.

Finally, we note that estimation of Fisher information falls under the umbrella of \emph{estimation of nonlinear functionals}; see for example \cite{birge1995estimation}.  Most of the commonly used information measures, such as entropy, relative entropy, and mutual information, are nonlinear functionals and their estimation has recently received considerable attention; the interested reader is referred to \cite{ sricharan2012estimation,wu2016minimax,han2017optimal,verdu2019empirical} and references therein.

The main contributions and the paper outline are as follows:
\begin{description}  
    \item[Section~\ref{sec:related_work}] is dedicated to a literature review of existing estimators of Fisher information;

    \item[Section~\ref{sec:bhattacharya_est}] revisits the Bhattacharya estimator. In particular, Theorem~\ref{thm:FIestError} provides  explicit and tighter non-asymptotic bounds on its convergence rate, improving the results in \cite{bhattacharya1967estimation} and \cite{dmitriev1974estimation}. Furthermore, Theorem~\ref{thm:modFIestError} provides an alternative bound for the Bhattacharya estimator under the additional assumption that the density function is upper bounded within any given interval. The explicit non-asymptotic results enable us to see that the Bhattacharya estimator needs an extremely large number of samples to guarantee a specific error within given confidence interval; 

    \item[Section~\ref{sec:ClippedEstimator}] proposes a new estimator, termed clipped estimator, which is designed to remedy the  large required sample size of Bhattacharya estimator. In particular, Theorem~\ref{thm:clippedFIestError} shows that the clipped estimator has better bounds on rates of convergence than the Bhattacharya estimator, albeit with different assumptions on the PDF; 

    \item[Section~\ref{sec:est_GaussianNoise}] evaluates the convergence rates of the two estimators for the practically relevant case of a random variable contaminated by Gaussian noise. Moreover, using Brown's identity, which relates the Fisher information and the minimum mean squared error (MMSE), consistent estimators for the MMSE are proposed and their rates of convergence  are evaluated in Proposition~\ref{prop:MMSEexample};

    \item[Section~\ref{sec:example}] is dedicated to simulation examples; and 

    \item[Section~\ref{sec:conclusion}] concludes the paper.
\end{description}

\paragraph*{Notation} Throughout the paper deterministic quantities are denoted by lowercase letters, and random variables are denoted by uppercase letters.  The expected value and variance of $X$ are denoted by $\E[X]$ and $\var(X)$, respectively.  The gamma function is denoted by $\Gamma(\cdot)$. Moreover, unless stated otherwise, $f_n$ and $\hat{f}$ denote the estimator of $f$, which is a random variable, and the corresponding estimate, which is a realization, respectively.

\section{Available Estimators}
\label{sec:related_work}

As aforementioned, the estimation of the Fisher information was first studied by Bhattacharya in \cite{bhattacharya1967estimation}. The Bhattacharya estimator is given by 
\begin{align}
    I_{n}
    = \int_{ |t| \le k_n} \frac{\left(f_n'(t)\right)^2}{f_n(t)} \dr t,
    \label{eq:FI_kernel_estimator}
\end{align}
where $k_n \ge 0$ determines the integration interval as a function of the sample size $n$ and the unknown functions $f$ and $f'$ are replaced by their kernel estimates, that is, 
\begin{align}
    f_n(t)
    &=\frac{1}{n} \sum_{i=1}^n \frac{1}{a_0} K \left( \frac{t-Y_i}{a_0}\right) \\
    f'_n(t)
    &=\frac{1}{n} \sum_{i=1}^n \frac{1}{a_1} K' \left( \frac{t-Y_i}{a_1}\right).
\end{align}
Here $a_0, a_1 > 0$ are bandwidth parameters, and $K(\cdot)$ denotes the kernel, which is assumed to satisfy certain regularity conditions. 

Let $F$ and $F_n$ denote a CDF and an empirical CDF respectively. Then, Donoho's estimator for Fisher information is given by \cite{donoho1988one}
\begin{align}
    I_\text{D}(\epsilon)
    = \inf \left\{I(G): \sup_t |G(t)-F_n(t)| \le \epsilon \right\},
    \label{eq:FI_donoho_estimator}
\end{align}
for some $\epsilon>0$. 
Donoho's estimator is based on the idea of one-sided confidence intervals. More precisely, while the true Fisher information of a density $f$ cannot be upper bounded based on samples alone, it can be lower bounded by minimizing Fisher information over a suitably chosen set of densities. Choosing the latter as a density ball, in terms of the Kolmogorov distance, centered at the empirical CDF establishes a connection to the observed samples. Donoho showed that with a computable probability this estimator provides a lower bound on the true Fisher information, and that the radius of the Kolmogorov distance ball can be reduced in such a way, as the sample number increases, that the estimator in \eqref{eq:FI_donoho_estimator} is consistent. In practice, however, Donoho's estimator requires solving a constrained variational optimization problem, whose structure is closely related the one considered by Huber in \cite{huber1974fisher}. The solution of this type of problem is given by a certain type of non-polynomial spline approximation. The corresponding fitting problem, however, is notoriously hard to solve numerically, even for small sample sizes, which often prevents the Donoho estimator from being useful in practice.

Among the available approaches for estimation of the Fisher information, the plug-in Bhattacharya estimator is the most straightforward and the easiest to implement. Therefore, a thorough understanding of the Bhattacharya estimator is of practical importance.   In \cite{bhattacharya1967estimation,dmitriev1974estimation} and \cite{nadaraya2016integral}, the authors focused on the asymptotic regime but did not consider the finite sample size case. More specifically, there are inexplicit constants in their bounds. In this work, we extend the analysis to the non-asymptotic regime. In addition, considering the potential real-time applications of the Fisher information/MMSE estimation (\textit{e.g.}, the implementation of the Mercury/waterfilling power allocation in terms of the MMSE \cite{lozano2006optimum}), the complexity of the estimation is also taken into account.   The Bhattacharya estimator is analyzed next.

\section{Bhattacharya Estimator}
\label{sec:bhattacharya_est}

In this section, we revisit the asymptotically consistent estimator proposed by Bhattacharya in \cite{bhattacharya1967estimation} and produce explicit and non-asymptotic bounds.

\subsection{Estimating Density and Its Derivatives} 

In order to analyze plug-in estimators it is necessary to obtain rates of convergence for $f_n$ and $f'_n$, that is, the kernel estimators of the density and its derivative. The following theorem, which is largely based on the proof by Schuster  in \cite{schuster1969estimation}, presents such rates. The proof in \cite{schuster1969estimation} makes use of the Dvoretzky-Kiefer-Wolfowitz (DKW) inequality for the empirical CDF. The next theorem refines the proof of \cite{schuster1969estimation} by using the best possible constant for the DKW inequality shown in \cite{massart1990tight}.  

\begin{theorem}\label{thm:boundsEstDensityandDerivatives}
Let $r \in \{0,1\}$ and 
\begin{align}
    v_r 
    &= \int \bigl| k^{(r+1)}(t) \bigr| \dr t,
    \label{eq:def_vr}\\
    \delta_{r,a} 
    &= \sup_{t \in \mathbb{R}} \left| \E\left[f_n^{(r)}(t)\right]- f^{(r)}(t)  \right|. 
    \label{eq:def_deltara}
\end{align}
Then, for any $\epsilon> \delta_{r,a}$ and any $n \geq 1$ the following bound holds: 
\begin{align}
    \mathbb{P} \left[ \sup_{t \in \mathbb{R}} \left| f_n^{(r)}(t)-f^{(r)}(t) \right| >\epsilon  \right] 
    \le 2 {\rm e}^{-2n \frac{a^{2r+2} (\epsilon- \delta_{r,a})^2}{ v_r^2} }. 
\end{align} 
\end{theorem}

\begin{IEEEproof}
See Appendix~\ref{app:thm_boundsEstDensityandDerivatives}. 
\end{IEEEproof}

\subsection{Analysis of the Bhattacharya Estimator} 

The following theorem is a non-asymptotic refinement of the  result obtained by Bhattacharya in \cite[Theorem~3]{bhattacharya1967estimation} and Dmitriev and  Tarasenko in \cite[Theorem~1]{dmitriev1974estimation}.

\begin{theorem}\label{thm:FIestError}
Assume there exists a function $\phi$ such that 
\begin{align}
\sup_{|t| \le x} \frac{1}{f(t)} \le \phi(x)   \text{ for all  $x$}. \label{eq:Definition of phi}
\end{align} 
Then, provided that
\begin{align}
   \sup_{|t| \le k_n} \left|f_n^{(r)}(t)-  f^{(r)}(t) \right|
   & \le \epsilon_r,  \, r \in \{0,1\},
   \label{eq:epsilon_i} 
\end{align}
and 
\begin{align}
    \epsilon_0 \phi(k_n) < 1 , 
    \label{eq:epsilon_phi}
\end{align}
the following bound holds: 
\begin{align}
    \left |I(f) -  I_{n} \right| 
    &\le \frac{4  \epsilon_1 k_n \rho_{\max}(k_n)+ 2 \epsilon_1^2  k_n \phi(k_n) + \epsilon_0\phi(k_n) I(f) }{1 -\epsilon_0 \phi(k_n)} + c(k_n), 
    \label{eq:boundDiffFI}
\end{align}
where 
\begin{align}
    \rho_{\max}(k_n)
    &= \sup_{|t| \le k_n} \left| \frac{f'(t)}{f(t)} \right|, 
    \label{eq:rhomax}\\
    c(k_n)
    &= \int_{|t| \ge k_n}  \frac{(f'(t))^2}{f(t)} \dr t . 
    \label{eq:c_kn}
\end{align}
\end{theorem}

\begin{IEEEproof}
See Appendix~\ref{app:thm_FIestError}.
\end{IEEEproof} 

The bound in \eqref{eq:boundDiffFI} is an improvement of the original bound in \cite{bhattacharya1967estimation} and \cite{dmitriev1974estimation}, which contains terms of the form $\epsilon_0 \phi^4(k_n)$.  

Note that $\phi(k_n)$ in \eqref{eq:Definition of phi} can be rapidly increasing with $k_n$. For example, as will be shown later, $\phi(k_n)$ increases super-exponentially with $k_n$ for a random variable contaminated by Gaussian noise. This implies that, while the Bhattacharya estimator converges, the rate of convergence guaranteed by the bound in \eqref{eq:boundDiffFI} is extremely slow. 
A modified bound is proposed in the subsequent theorem.

\begin{theorem}\label{thm:modFIestError}
Assume that $f(t)$ is bounded on the interval $t \in [-k_n,k_n]$, \textit{i.e.},
\begin{align}
    \sup_{|t| \le k_n } f(t) \le f_0. \label{eq:sup_f}
\end{align}
If the assumptions in \eqref{eq:Definition of phi}, \eqref{eq:epsilon_i}, and \eqref{eq:epsilon_phi} hold, 
then
\begin{align}
    | I(f) - I_n | 
    &\le \big(  \epsilon_1 \left( 4 + d_f(k_n) + d_{f_n}(k_n) \right) + \epsilon_0 \left( 2+ d_{f_n}(k_n) \right)  \rho_{\max}(k_n) \big) \psi(\epsilon_0,k_n)  + c(k_n).
    \label{eq:mod_diffFI}
\end{align}
where $\rho_{\max}$ and $c$ are given by \eqref{eq:rhomax} and \eqref{eq:c_kn}, respectively,
\begin{equation}
    \psi(\epsilon_0,k_n) = \max \left( \log(f_0 + \epsilon_0), \log\left(\frac{\phi(k_n)}{1-\epsilon_0 \phi(k_n)} \right) \right),
\label{eq:def_psi}
\end{equation}
and $d_g(k_n)$ denotes the number of zeros of the derivative of the function $g$ on the interval $[-k_n,k_n]$, \textit{i.e.},
\begin{equation}
    d_g(k_n)
    = \left\lvert \left\{ x \in [-k_n,k_n]: g'(x)=0  \right\}  \right\rvert.
    \label{eq:def_d}
\end{equation}
\end{theorem}

\begin{IEEEproof}
See Appendix~\ref{app:thm_modFIestError}.
\end{IEEEproof}

\begin{remark}
Note that $\psi$ in \eqref{eq:mod_diffFI} is on the order of $\log(\phi(k_n))$ which typically increases much slower with $k_n$ than $\phi$ in \eqref{eq:boundDiffFI}. As a result, the bound in Theorem~\ref{thm:modFIestError} can lead to a better bound on the convergence rate than that in Theorem~\ref{thm:FIestError}, given appropriate upper bounds on $d_{f}$ and $d_{f_n}$.  
Since Gaussian blurring of the original 1-dimensional function never creates new maxima, we have that $d_{f_Y} \le d_{f_X}$, which is a constant. 
However, to the best of our knowledge, the only known upper bound on $d_{f_n}$ is given by $d_{f_n} \le n$ \cite[Theorem~2]{carreira2003number}, which is not useful in practice. 
Despite this drawback, we decided to include Theorem~\ref{thm:modFIestError} for completeness and in the hope that tighter bounds on $d_{f_n}$ might be established in the future.
\end{remark}

The main problem in the convergence analysis of the estimator in \eqref{eq:FI_kernel_estimator} is that $1/f_n(x)$ is only bounded if $f(x) > \epsilon_0$. For distributions with sub-Gaussian tails, this implies that the interval $[-k_n, k_n]$, on which this is guaranteed to be the case, grows sub-logarithmically (compare Theorem~\ref{thm:FIestError_GaussianNoise}), causing the required number of  samples to grow super-exponentially. In next section, we propose an estimator that has better  guaranteed rates of convergence.

\section{A Clipped Estimator}
\label{sec:ClippedEstimator}

In order to remedy the slow  guaranteed convergence rates of the Bhattacharya estimator, we dispense with the tail assumption in \eqref{eq:Definition of phi}, but introduce the new assumption that the unknown true score function $\rho(t) = f'(t)/f(t)$ is bounded (in absolute value) by a known function $\overline{\rho} $. This allows us to clip $f'_n(x)/f_n(x)$ and in turn $1/f_n(x)$ without affecting the consistency of the estimator. 

\begin{theorem}\label{thm:clippedFIestError}
Assume there exists a function $\overline{\rho} $ such that 
\begin{align}
    \lvert \rho(t) \rvert  \leq \lvert \overline{\rho}(t) \rvert,
\end{align} 
for all $t \in \mathbb{R}$ 
and let
\begin{equation}
    I_n^\text{c}
    = \int_{-k_n}^{k_n} \min\left\{ \lvert \rho_n(t) \rvert \,,\, \lvert \overline{\rho}(t) \rvert \right\} \, \lvert f'_n(t) \rvert \,\dr t,
    \label{eq:clipped_estimator}
\end{equation}
where 
\begin{equation}
    \rho_n(t) = \frac{f_n'(t)}{f_n(t)}.
\end{equation}
Under the assumptions in \eqref{eq:epsilon_i}, it holds that 
\begin{align}
    \lvert I(f) - I_n^\text{c} \rvert 
    &\leq \max\Big\{ 4 \epsilon_1 \Phi^1(k_n) + 2 \epsilon_0 \Phi^2(k_n)  + c(k_n), 3 \epsilon_1 \Phi_\text{max}^1(k_n)  + \epsilon_0 \Phi_\text{max}^2(k_n) \Big\} \\
    &\leq 4 \epsilon_1 \Phi_\text{max}^1(k_n)  + 2 \epsilon_0 \Phi_\text{max}^2(k_n)  + c(k_n),
    \label{eq:diffFI_clipped}
\end{align}
where $c(k_n)$ is defined in \eqref{eq:c_kn} and
\begin{align}
    \Phi^m(x) 
    &= \int_{-x}^x \left\lvert \rho^m(t) \right\rvert\, \dr t, \\
    \Phi_\text{max}^m(x) 
    &= \int_{-x}^x \left\lvert \overline{\rho} ^m(t) \right\rvert \, \dr t.
\end{align}
In addition, if $f(t)$ is bounded as in \eqref{eq:sup_f}, then
\begin{align}
    \Phi^m (k_n)
    &\leq \min\Big\{ (2 + d_f) \overline{\rho} ^{m-1}(k_n) \psi(0,k_n) \,,\, \Phi_\text{max}^m(k_n) \Big\},
    \label{eq:diffFI_clipped_mod}
\end{align}
where $\psi$ and $d_f$ are defined in \eqref{eq:def_psi} and \eqref{eq:def_d}, respectively.
\end{theorem}

\begin{IEEEproof}
See Appendix~\ref{app:thm_clippedFIestError}.
\end{IEEEproof} 

Note that, we can set $\overline{\rho}(k_n) = \rho_{\max}(k_n)$. Although $\rho_{\max}(k_n)$ also increases with $k_n$, it usually increases much slower than $\phi(k_n)$. For example, as shown later, $\rho_{\max}(k_n)$ is linear in $k_n$ in the Gaussian noise case. As a result, better bounds on the convergence rate can be shown for the clipped estimator.

\section{Estimation of the Fisher Information of\\ a Random Variable Contaminated by Gaussian Noise}
\label{sec:est_GaussianNoise}

This section evaluates the results of Section~\ref{sec:bhattacharya_est} and Section~\ref{sec:ClippedEstimator} for the important special case of a random variable contaminated by Gaussian noise. To this end, we let $f_Y$ denote the PDF of a random variable 
\begin{align}
    Y = Y_{\snr} = \sqrt{\snr} X + Z,
\end{align}
where $\snr > 0$ is a signal-to-noise-ratio parameter, $X$ is an arbitrary random variable, $Z$ is a standard Gaussian random variable, and $X$ and $Z$ are independent.  We are interested in estimating the Fisher information of $f_Y$.  We only make the very mild assumption that $X$ has a finite second moment but otherwise it is allowed to be an arbitrary random variable. We also assume that $\snr$ is known.

The Fisher information of $f_Y$ is connected to other estimation and information measures via several important identities.  In particular, the Fisher information can be connected to the quadratic Bayesian risk or the MMSE as follows:
\begin{align}
     I(f_Y)& = 1-\snr \mmse(X|Y), 
     \label{eq:BrownIdentity}
\end{align}
where the MMSE  is given by 
\begin{align}
    \mmse(X|Y) = \mathbb{E}\left[(X-\E[X|Y])^2\right]. 
\end{align}
In the statistics literature, this relationship is known as Brown's identity  \cite{brown1971admissible}. The Fisher information can also be connected to information measures such as mutual information,  entropy, and continuous entropy  via the following identities:
\begin{align}
    2 I(X; Y_{\snr})
    &= \int_0^{\snr}  \mmse(X|Y_\gamma) \dr \gamma 
    = \int_0^{\snr} \frac{1- I(f_{Y_{\gamma}})}{\gamma} \dr \gamma, \label{eq:I_MMSE}\\
    2H(X)
    &= \int_0^\infty  \mmse(X|Y_{\gamma}) \dr \gamma 
    = \int_0^\infty  \frac{1- I(f_{Y_{\gamma}})}{\gamma} \dr \gamma, 
    \label{eq:EntropyDiscrete} \\
    2 h(X)
    &= \int_0^\infty \frac{1- I(f_{Y_{\gamma}})}{\gamma} -\frac{1}{2 \pi {\rm e} + \gamma} \dr \gamma. \label{eq:EntropyContinious}
\end{align}
The relationship in \eqref{eq:I_MMSE} is known as the I-MMSE identity and was shown in \cite{guo2005mutual} together with the identity in \eqref{eq:EntropyDiscrete}.   The identity in \eqref{eq:EntropyContinious} is known as De Bruijin's identity and holds if $\lim_{\snr \to \infty}h \left(X+\frac{1}{\sqrt{\snr}}Z\right)=h(X)$. It was show in \cite{stam1959some}; see also \cite{guo2005mutual} for an alternative proof.

Using the estimator of the Fisher information together with the above identities it should be possible to construct estimators for mutual information, entropy, and continuous entropy. In what follows, we will use the identity in \eqref{eq:BrownIdentity} to  propose an estimator for the MMSE and will evaluate the performance of that estimator.   We note that the idea of using the I-MMSE identity in \eqref{eq:I_MMSE} to estimate the mutual information has been already used in \cite{alghamdi2019mutual}. Note, however, that the approach in \cite{alghamdi2019mutual} requires the existence of all moments of the distribution of $X$, while here we only require the existence of the second moment.  

The following lemma provides explicit expressions for the quantities appearing in Section~\ref{sec:bhattacharya_est} and Section~\ref{sec:ClippedEstimator} that are needed to evaluate the error bounds  for the Bhattacharya and the clipped estimator. 

\begin{lemma} \label{lem:FIestErrGaussianNoise} 
Let $K(t)=\frac{1}{\sqrt{2 \pi}} {\rm e}^{-\frac{t^2}{2}}$. Then,
\begin{align}
    \delta_{r,a}
    &= a \cdot \begin{cases}   
        \frac{1}{\sqrt{2 \pi }} \frac{1}{\sqrt{\rm e}}, & r=0  \\[1.5ex]
        \frac{\frac{2}{{\rm e}}+1}{\sqrt{2 \pi }},  & r=1, 
    \end{cases}\\
    v_r
    &= \begin{cases}
        \sqrt{\frac{2}{\pi}},  & r=0\\
        \sqrt{\frac{2}{ {\rm e}\pi}}, & r=1,
    \end{cases} \\
    \rho_{\max}(k_n) &\le \sqrt{3 \snr \var(X) } +3 k_n , \\
     I(f_Y) &\le 1,\\
     \phi(t) &\le  \sqrt{2 \pi}{\rm e}^{(t^2 + \snr \E[X^2])}.
\end{align}
\end{lemma} 

\begin{IEEEproof}
See Appendix~\ref{prof:lem:FIestErrGaussianNoise}. 
\end{IEEEproof}

We now bound $ c (k_n)$. To this end, we need the notion of sub-Gaussian random variables: a random variable $X$ is said to be $\alpha$-sub-Gaussian if 
\begin{align}
    \E[{\rm e}^{t X}] \le {\rm e}^{\frac{\alpha^2 t^2}{2}}, \forall t\in \mathbb{R}. 
\end{align}

\begin{lemma} \label{lem:ckn}
Suppose that $\E[X^2]<\infty$. Then, 
\begin{align}
    c (k_n)  \le  \inf_{ v >0} \frac{2 \Gamma^{ \frac{1}{(1+v)}} \left(v+\frac{1}{2} \right)}{ \pi^{ \frac{1}{2(1+v)}}} \left(  \frac{ \snr \E[|X|^2] +1}{k_n^2} \right)^{\frac{v}{1+v}}.
\end{align}
In addition, if $|X|$ is $\alpha$-sub-Gaussian, then
\begin{align}
    c (k_n)  
    &\le \inf_{ v >0} \frac{2 \Gamma^{ \frac{1}{(1+v)}} \left(v+\frac{1}{2} \right)}{ \pi^{ \frac{1}{2(1+v)}}} \left( 2 {\rm e}^\frac{\alpha^2 \snr - k_n^2}{2} \right)^{ \frac{v}{1+v} }. 
\end{align}
\end{lemma}

\begin{IEEEproof}
See Appendix~\ref{app:lem_ckn}. 
\end{IEEEproof}

\subsection{Convergence of the Bhattacharya Estimator}

By combining the results in Theorem~\ref{thm:boundsEstDensityandDerivatives}, Theorem~\ref{thm:FIestError}, Lemma~\ref{lem:FIestErrGaussianNoise}, and Lemma~\ref{lem:ckn} we have the following theorem. 

\begin{theorem}\label{thm:FIestError_GaussianNoise}
Let $K(t)=\frac{1}{\sqrt{2 \pi}} {\rm e}^{-\frac{t^2}{2}}$. If $a = n^{-w}$, where $w\in\left(0,\frac{1}{6} \right)$, and $k_n = \sqrt{ u \log(n) }$, where $u \in \left(0,w \right)$, then
\begin{align}
    \mathbb{P} \left[ \left| I_n - I(f_Y) \right| \ge \varepsilon_n \right] 
    &\le 2 {\rm e}^{-c_1 n^{1-4w} } + 2 {\rm e}^{-c_2 n^{1-6 w} }, 
    \label{eq:boundProbDiffFI} 
\end{align}
where 
\begin{align}
    \varepsilon_n 
    &\le \frac{ n^{-w} \sqrt{u \log(n)} \left( 4 c_3 + 12 \sqrt{u \log(n)} + 2 c_5 n^{u-w} \right) }{1 - n^{u-w}} + \frac{c_4}{\sqrt{ u \log(n) }} + \frac{ c_5 }{n^{w-u} - 1}, 
    \label{eq:boundvarepsilon}
\end{align}
and where the constants are given by
\begin{align}
    c_1 &= \pi \left(1 - \frac{1}{\sqrt{2 \pi {\rm e}}}  \right)^2, \\
    c_2& = {\rm e} \pi \left(1 -  \frac{\frac{2}{{\rm e}}+1}{\sqrt{2 \pi }} \right)^2,\\
    c_3 &= \sqrt{3 \snr \var(X)}, \\
    c_4 &=  \frac{2 \Gamma^{ \frac{1}{2}} \left(\frac{3}{2} \right) \sqrt{ \snr \E[|X|^2] +1 }}{ \pi^{ \frac{1}{4}}} , \\
    c_5 &= \sqrt{2 \pi} {\rm e}^{\snr \E\left[X^2\right]} .
\end{align}
In addition, if $|X|$ is $\alpha$-sub-Gaussian, then
\begin{align}
    \varepsilon_n 
    &\le \frac{ n^{-w} \sqrt{u \log(n)} \left( c_3 + 12 \sqrt{u \log(n)} + 2 c_5 n^{u-w} \right) }{1 - n^{u-w}} + \frac{ c_5 }{n^{w-u} - 1} + c_6 n^{-\frac{u}{4}}, 
    \label{eq:FIerrBound_subGaussian}
\end{align}
where
\begin{align}
    c_6 &= \frac{2^{\frac{3}{2}} \Gamma^{ \frac{1}{2}} \left(\frac{3}{2} \right) {\rm e}^{\frac{\alpha^2 \snr}{4}} }{ \pi^{ \frac{1}{4}}} .
    \label{eq:c6}
\end{align}
\end{theorem}

\begin{IEEEproof}
See Appendix~\ref{app:thm_FIerr_GaussianNoise}.
\end{IEEEproof}


The choice of $u$ and $w$ results in a trade-off between precision, $\varepsilon_n$, and confidence, \textit{i.e.} the probability of the estimation error exceeding $\varepsilon_n$. On the one hand, small values of $u$ and large values of $w$ result in better precision at the cost of a lower confidence. On the other hand, large values of $u$ and small values of $w$ improve the confidence but deteriorate the precision. In turn, this also affects the convergence rates, meaning that faster convergence of the precision can be achieved at the expense of a slower convergence of the confidence and vice versa.

\subsection{Convergence of the Clipped Estimator}

From the  evaluation of the Bhattacharya estimator in Theorem~\ref{thm:FIestError_GaussianNoise}, it is apparent that the bottleneck term is the truncation parameter $k_n = \sqrt{ u \log(n) }$, which results in slow precision decay of the order $ \varepsilon_n=O \left( \frac{1}{\sqrt{ u \log(n) }} \right)$.  Next, it is shown that the clipped estimator results in an improved precision over the Bhattacharya estimator.  Specifically, the precision will be shown to decay polynomially in $n$ instead of logarithmically. 

By utilizing the results in Theorem~\ref{thm:boundsEstDensityandDerivatives}, Lemma~\ref{lem:FIestErrGaussianNoise}, and Lemma~\ref{lem:ckn}, we specialize the result in Theorem~\ref{thm:modFIestError} to the Gaussian noise case.

\begin{theorem}\label{thm:clipped_FIest_GaussianNoise}
Let $K(t)=\frac{1}{\sqrt{2 \pi}} {\rm e}^{-\frac{t^2}{2}}$. If $a_0 = n^{-w_0}$, where $w\in\left(0,\frac{1}{4} \right)$, $a_1 = n^{-w}$, where $w_1\in\left(0,\frac{1}{6} \right)$, and $k_n = n^u$, where $u \in \left(0, \min \left(\frac{w_0}{3},\frac{w_1}{2}\right) \right)$, then 
\begin{align}
    \mathbb{P} \left[ \left| I_n^\text{c} - I(f_Y)  \right| \ge \varepsilon_n \right] 
    &\le 2 {\rm e}^{-c_1 n^{1-4 w_0}} + 2 {\rm e}^{-c_2 n^{1-6 w_1}}, 
\end{align}
where 
\begin{align}
    \varepsilon_n 
    &\le 4 n^{3 u-w_0}\left( c_3 n^{-2u} + 3 n^{-u} + 3 \right) + 4 n^{2 u-w_1}\left( 2 c_3 n^{-u} + 3 \right) + c_4 n^{-u},
    \label{eq:boundnewFIerr}
\end{align}
and the constants $c_i , i\in [1:4]$ are as in Theorem~\ref{thm:FIestError_GaussianNoise}.
In addition, 
if $|X|$ is $\alpha$-sub-Gaussian, then
\begin{align}
    \varepsilon_n 
    &\le 4 n^{3 u-w_0}\left( c_3 n^{-2u} + 3 n^{-u} + 3 \right) + 4 n^{2 u-w_1}\left( 2 c_3 n^{-u} + 3 \right) + c_6 {\rm e}^{-\frac{n^{2u}}{4}}, 
    \label{eq:clippedFIerrBound_subGaussian}
\end{align}
where $c_6$ is given by \eqref{eq:c6}.
\end{theorem}

\begin{IEEEproof}
See Appendix~\ref{app:thm_clippedFIest}.
\end{IEEEproof}

\subsection{Applications to the Estimations of the MMSE}

Using Brown's identity in \eqref{eq:BrownIdentity}, we propose the following estimators for the MMSE: 
\begin{align}
    \mmse_n(X,\snr) = \frac{1-  I_n}{\snr},
\end{align}
and
\begin{align}
    \mmse_n^\text{c}(X,\snr) = \frac{1-  I_n^\text{c}}{\snr}.
\end{align}
The results for the estimators of Fisher information in Theorem~\ref{thm:FIestError_GaussianNoise} and Theorem~\ref{thm:clipped_FIest_GaussianNoise} can be immediately extended to the MMSE estimators as follows.

\begin{proposition}\label{prop:MMSEexample}
Let $K(t)=\frac{1}{\sqrt{2 \pi}} {\rm e}^{-\frac{t^2}{2}}$. If $a = n^{-w}$, where $w\in\left(0,\frac{1}{6} \right)$, and $k_n = \sqrt{ u \log(n) }$, where $u \in \left(0,w \right)$, then
\begin{align}
    \mathbb{P} \left[ \left| \mmse_n(X,\snr) - \mmse(X,\snr) \right| \ge \snr \varepsilon_n \right] 
    &\le 2 {\rm e}^{-c_1 n^{1-4w} } + 2 {\rm e}^{-c_2 n^{1-6w} },
\end{align}
where $\varepsilon_n$, $c_1$, and $c_2$ are given by Theorem~\ref{thm:FIestError_GaussianNoise}. 
\end{proposition}

\begin{proposition}\label{prop:MMSEexample_clipped}
Let $K(t)=\frac{1}{\sqrt{2 \pi}} {\rm e}^{-\frac{t^2}{2}}$. If $a_0 = n^{-w_0}$, where $w\in\left(0,\frac{1}{4} \right)$, $a_1 = n^{-w}$, where $w_1\in\left(0,\frac{1}{6} \right)$, and $k_n = n^u$, where $u \in \left(0, \min \left(\frac{w_0}{3},\frac{w_1}{2}\right) \right)$, then 
\begin{align}
    \mathbb{P} \left[ \left| \mmse_n^\text{c}(X|Y) - \mmse(X|Y) \right| \ge \snr \varepsilon_n \right] 
    &\le 2 {\rm e}^{-c_1 n^{1-4 w_0}} + 2 {\rm e}^{-c_2 n^{1-6 w_1}}
\end{align}
where $\varepsilon_n$, $c_1$, and $c_2$ are given by Theorem~\ref{thm:clipped_FIest_GaussianNoise}. 
\end{proposition}

\section{Examples}
\label{sec:example}

This section provides numerical and simulation results to demonstrate the performance of the estimators. We focus on the setup considered in Section~\ref{sec:est_GaussianNoise} where $Y$ is a random variable contaminated by Gaussian noise. First, we present and compare representative examples of the estimates of interest, including the density function, its derivative, the Fisher information, and the MMSE. Second, the bias and variance of the proposed estimators are demonstrated. Finally, the sample complexities (\textit{i.e.}, the number of samples needed to guarantee a given precision and a given confidence) of the proposed estimators are compared. MATLAB codes for all simulations can be found in \cite{cao2020estimationcode}.

In each experiment, we examine the estimators in the following two example scenarios: 1) a continuous example in which the input distribution is a standard Gaussian distribution; 2) a non-continuous example in which the input distribution is binary such that $X=1$ with probability 0.5 and $X=-1$ with probability 0.5. One reason for choosing these two cases as examples is that closed-form expressions for the Fisher information and the MMSE exist for both.

Moreover, both the standard Gaussian input and the binary input are $\alpha$-sub-Gaussian. More specifically, for the standard Gaussian input,
\begin{align}
    \E \left[ {\rm e}^{t X} \right]
    &= \int_{x \in \mathbb{R}} {\rm e}^{t x} \frac{1}{\sqrt{2 \pi}} {\rm e}^{- \frac{x^2}{2}} \dr x 
    = {\rm e}^{\frac{t^2}{2}}.
\end{align}
Meanwhile, for the binary input,
\begin{align}
    \E \left[ {\rm e}^{t X} \right]
    &= \frac{{\rm e}^t + {\rm e}^{-t}}{2} \\
    &\le {\rm e}^{\frac{t^2}{2} \sup_{t\in \mathbb{R}} \frac{2}{t^2} \log\left( \frac{{\rm e}^t + {\rm e}^{-t}}{2} \right) }\\
    &= {\rm e}^{\frac{t^2}{2}}.
\end{align}
Therefore, both the standard Gaussian distribution and binary distribution are sub-Gaussian with proxy variance $\alpha = 1$.

\begin{figure*}[!t]
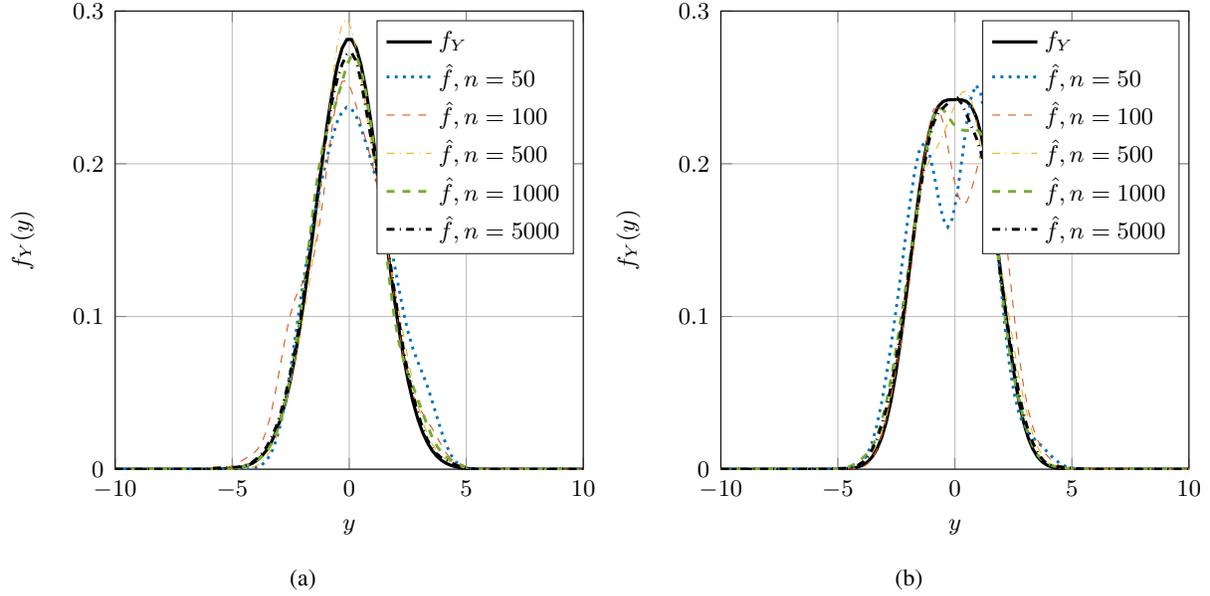

\centering{
\subfloat[]{\input{f_G.tex}}
\subfloat[]{\input{f_B.tex}}
}
\caption{Comparison of the density function and the density  estimates with: a) Gaussian input; and b) binary input.}
\label{fig:DensityEst}
\end{figure*}

\begin{figure*}[!t]
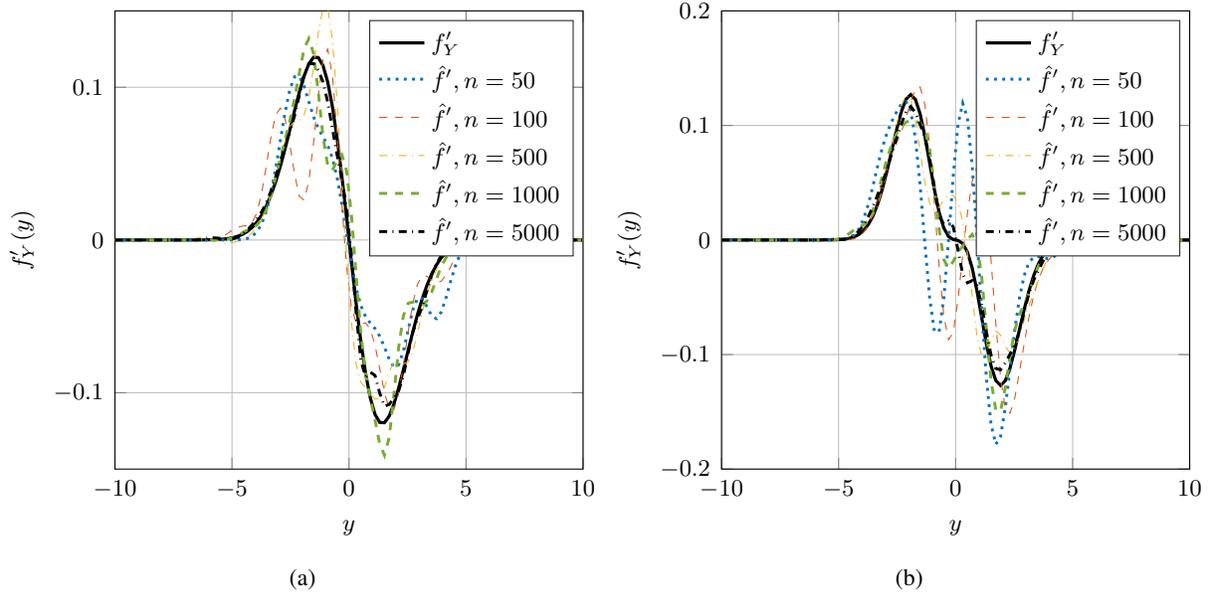

\centering{
\subfloat[]{\input{df_G.tex}}
\subfloat[]{\input{df_B.tex}}
}
\caption{Comparison of the density derivative and the derivative of the density  estimates with: a) Gaussian; and b) binary input.}
\label{fig:DerivativeEst}
\end{figure*}

\begin{figure*}[!t]
\centering{
\subfloat[]{
%
%
\definecolor{mycolor1}{rgb}{0.00000,0.44700,0.74100}%
\begin{tikzpicture}
\pgfplotsset{every tick label/.append style={font=\footnotesize}}

\begin{axis}[%
width=2.45in,
height=2.4in,
at={(0.758in,0.481in)},
scale only axis,
xmin=0,
xmax=10,
xlabel style={font=\color{white!15!black},font=\footnotesize},
xlabel={$\snr$},
ymin=0,
ymax=1,
ylabel style={font=\color{white!15!black},font=\footnotesize},
ylabel={$I(f_Y)$},
axis background/.style={fill=white},
xmajorgrids,
ymajorgrids,
legend style={legend cell align=left, align=left, draw=white!15!black,font=\footnotesize}
]
\addplot [color=black, line width=1pt, mark options={solid, black}]
  table[row sep=crcr]{%
0	1.00000011520611\\
0.416666666666667	0.705882273059894\\
0.833333333333333	0.545455336733916\\
1.25	0.444442070717819\\
1.66666666666667	0.374999428564873\\
2.08333333333333	0.324329121135663\\
2.5	0.285719529912664\\
2.91666666666667	0.255318179435321\\
3.33333333333333	0.230760303141204\\
3.75	0.210513337754681\\
4.16666666666667	0.193537823263802\\
4.58333333333333	0.179104424411649\\
5	0.166666691974692\\
5.41666666666667	0.155844275802255\\
5.83333333333333	0.146341652661674\\
6.25	0.137957867028584\\
6.66666666666667	0.130458556231506\\
7.08333333333333	0.123727639798906\\
7.5	0.117652675226519\\
7.91666666666667	0.112142736194137\\
8.33333333333333	0.107142376800673\\
8.75	0.10256359679969\\
9.16666666666667	0.0983602188840743\\
9.58333333333333	0.0944879090974997\\
10	0.0909090416313405\\
};
\addlegendentry{$I(f_Y)$}

\addplot [color=black, dashed, mark=x, mark options={solid, black}]
  table[row sep=crcr]{%
0	0.747925766397812\\
0.416666666666667	0.575090819736449\\
0.833333333333333	0.465878765064369\\
1.25	0.391438009981003\\
1.66666666666667	0.337573466038741\\
2.08333333333333	0.296834824473055\\
2.5	0.264962089626176\\
2.91666666666667	0.239333198744802\\
3.33333333333333	0.218245270450082\\
3.75	0.20059019919195\\
4.16666666666667	0.185620652614982\\
4.58333333333333	0.172755896447816\\
5	0.161595337962427\\
5.41666666666667	0.15183522775338\\
5.83333333333333	0.143225253108808\\
6.25	0.135548860068479\\
6.66666666666667	0.128621063901808\\
7.08333333333333	0.122330386558629\\
7.5	0.116601793830618\\
7.91666666666667	0.111342965365872\\
8.33333333333333	0.106459412181396\\
8.75	0.101881521799667\\
9.16666666666667	0.0975645860560928\\
9.58333333333333	0.0934778195987593\\
10	0.0895965560601711\\
};
\addlegendentry{$\hat{I}, a_0 = a_1 = 0.6$}

\addplot [color=mycolor1, dashed, mark=x, mark options={solid, mycolor1}, line width=0.8pt]
  table[row sep=crcr]{%
0	0.925286132564937\\
0.416666666666667	0.660612331699901\\
0.833333333333333	0.520778671995715\\
1.25	0.431342706653828\\
1.66666666666667	0.36992802259104\\
2.08333333333333	0.323637411193896\\
2.5	0.287474740386361\\
2.91666666666667	0.258959472969623\\
3.33333333333333	0.235882067046959\\
3.75	0.216394027720108\\
4.16666666666667	0.199549208495921\\
4.58333333333333	0.184995204484953\\
5	0.172327264984604\\
5.41666666666667	0.161249655083926\\
5.83333333333333	0.151816106092481\\
6.25	0.143415085863205\\
6.66666666666667	0.135633496187691\\
7.08333333333333	0.128537274798331\\
7.5	0.122579715245913\\
7.91666666666667	0.117657717819441\\
8.33333333333333	0.113125303671689\\
8.75	0.108868609537505\\
9.16666666666667	0.104284143444466\\
9.58333333333333	0.09972174178723\\
10	0.0954835269791688\\
};
\addlegendentry{$\hat{I}, a_0 = a_1 = 0.3$}

\addplot [color=black, dashdotted, mark=o, mark options={solid, black}]
  table[row sep=crcr]{%
0	0.747925766397812\\
0.416666666666667	0.575090819736449\\
0.833333333333333	0.465878765064369\\
1.25	0.391438009981003\\
1.66666666666667	0.337573466038741\\
2.08333333333333	0.296834824473055\\
2.5	0.264962089626176\\
2.91666666666667	0.239333198744802\\
3.33333333333333	0.218245270450082\\
3.75	0.20059019919195\\
4.16666666666667	0.185620652614982\\
4.58333333333333	0.172755896447816\\
5	0.161595337962427\\
5.41666666666667	0.15183522775338\\
5.83333333333333	0.143225253108807\\
6.25	0.135548860068479\\
6.66666666666667	0.128621063901808\\
7.08333333333333	0.122330386558629\\
7.5	0.116601793830618\\
7.91666666666667	0.111342965365872\\
8.33333333333333	0.106459412181396\\
8.75	0.101881521799667\\
9.16666666666667	0.0975645860560928\\
9.58333333333333	0.0934778195987593\\
10	0.0895965560601711\\
};
\addlegendentry{$\hat{I}^\text{c}, a_0 = a_1 = 0.6$}

\addplot [color=mycolor1, dashdotted, mark=o, mark options={solid, mycolor1}, line width=0.8pt]
  table[row sep=crcr]{%
0	0.925286132547535\\
0.416666666666667	0.660612331688989\\
0.833333333333333	0.520778671995618\\
1.25	0.431342706653812\\
1.66666666666667	0.369928022591038\\
2.08333333333333	0.323637411193896\\
2.5	0.287474740386361\\
2.91666666666667	0.258959472969623\\
3.33333333333333	0.235882067046959\\
3.75	0.216394027720108\\
4.16666666666667	0.199549208495921\\
4.58333333333333	0.184995204484953\\
5	0.172327264984604\\
5.41666666666667	0.161249655083926\\
5.83333333333333	0.151816106092481\\
6.25	0.143415085863205\\
6.66666666666667	0.135633496187691\\
7.08333333333333	0.128537274798331\\
7.5	0.122579715245913\\
7.91666666666667	0.117657717819441\\
8.33333333333333	0.113125303671689\\
8.75	0.108868609537505\\
9.16666666666667	0.104284143444466\\
9.58333333333333	0.09972174178723\\
10	0.0954835269791688\\
};
\addlegendentry{$\hat{I}^\text{c}, a_0 = a_1 = 0.3$}

\end{axis}
\end{tikzpicture}%
\label{fig:FI_G}}
\subfloat[]{
%
%
\definecolor{mycolor1}{rgb}{0.00000,0.44700,0.74100}%
\begin{tikzpicture}
\pgfplotsset{every tick label/.append style={font=\footnotesize}}

\begin{axis}[%
width=2.45in,
height=2.4in,
at={(0.758in,0.481in)},
scale only axis,
xmin=0,
xmax=10,
xlabel style={font=\color{white!15!black},font=\footnotesize},
xlabel={$\snr$},
ymin=0.4,
ymax=1,
ylabel style={font=\color{white!15!black},font=\footnotesize},
ylabel={$I(f_Y)$},
axis background/.style={fill=white},
xmajorgrids,
ymajorgrids,
legend style={at={(0.66,0.02)}, anchor=south, legend cell align=left, align=left, draw=white!15!black,font=\footnotesize}
]
\addplot [color=black, line width=1pt, mark options={solid, black}]
  table[row sep=crcr]{%
0	1.00000011966156\\
0.416666666666667	0.710830109353098\\
0.833333333333333	0.578027408133109\\
1.25	0.527298027449293\\
1.66666666666667	0.522585174305776\\
2.08333333333333	0.543543393114816\\
2.5	0.57801743282852\\
2.91666666666667	0.618622636962733\\
3.33333333333333	0.660860504672101\\
3.75	0.70205683974922\\
4.16666666666667	0.740697597258397\\
4.58333333333333	0.776118390959039\\
5	0.807799456757379\\
5.41666666666667	0.835813946838782\\
5.83333333333333	0.860250196001293\\
6.25	0.881550900969542\\
6.66666666666667	0.899916858457115\\
7.08333333333333	0.915663678344477\\
7.5	0.929101455179423\\
7.91666666666667	0.940523051594606\\
8.33333333333333	0.95019772659591\\
8.75	0.958368331708985\\
9.16666666666667	0.965250779707787\\
9.58333333333333	0.971034900164989\\
10	0.975886103024781\\
};
\addlegendentry{$I(f_Y)$}

\addplot [color=black, dashed, mark=x, mark options={solid, black}]
  table[row sep=crcr]{%
0	0.930007429707687\\
0.416666666666667	0.671166819309514\\
0.833333333333333	0.54638387028819\\
1.25	0.493662841159199\\
1.66666666666667	0.482044700253026\\
2.08333333333333	0.49362771255568\\
2.5	0.518200418491903\\
2.91666666666667	0.54974213775037\\
3.33333333333333	0.584678075607306\\
3.75	0.62044523929558\\
4.16666666666667	0.655174492845167\\
4.58333333333333	0.687826034243776\\
5	0.718038744690074\\
5.41666666666667	0.745764394926241\\
5.83333333333333	0.771015579546663\\
6.25	0.793817274696957\\
6.66666666666667	0.814242614817971\\
7.08333333333333	0.832421019240936\\
7.5	0.848504827073016\\
7.91666666666667	0.862634546958263\\
8.33333333333333	0.874932676719704\\
8.75	0.885522987444145\\
9.16666666666667	0.894551986294129\\
9.58333333333333	0.902192104299868\\
10	0.908623179078866\\
};
\addlegendentry{$\hat{I}, a_0 = a_1 = 0.3$}

\addplot [color=mycolor1, dashed, mark=x, mark options={solid, mycolor1}, line width=0.8pt]
  table[row sep=crcr]{%
0	0.972025471259245\\
0.416666666666667	0.710429667058639\\
0.833333333333333	0.613294312999148\\
1.25	0.563068775792227\\
1.66666666666667	0.5551201575716\\
2.08333333333333	0.562434999824562\\
2.5	0.599427725591072\\
2.91666666666667	0.63063634760138\\
3.33333333333333	0.656586080413438\\
3.75	0.696994201555287\\
4.16666666666667	0.740799454485969\\
4.58333333333333	0.773443986196493\\
5	0.798151241474\\
5.41666666666667	0.823449829327684\\
5.83333333333333	0.85282558994858\\
6.25	0.883458077104579\\
6.66666666666667	0.907602568014685\\
7.08333333333333	0.924441641747133\\
7.5	0.9379114682134\\
7.91666666666667	0.949622821064049\\
8.33333333333333	0.959942899100426\\
8.75	0.969346805513485\\
9.16666666666667	0.97745927501845\\
9.58333333333333	0.983072413734669\\
10	0.985934430323197\\
};
\addlegendentry{$\hat{I}, a_0 = 0.3,  a_1 = 0.15$}

\addplot [color=black, dashdotted, mark=o, mark options={solid, black}]
  table[row sep=crcr]{%
0	0.930007429689084\\
0.416666666666667	0.671166819308507\\
0.833333333333333	0.546383870288144\\
1.25	0.493662841159183\\
1.66666666666667	0.482044700253022\\
2.08333333333333	0.493627712555679\\
2.5	0.518200418491903\\
2.91666666666667	0.54974213775037\\
3.33333333333333	0.584678075607306\\
3.75	0.62044523929558\\
4.16666666666667	0.655174492845167\\
4.58333333333333	0.687826034243776\\
5	0.718038744690074\\
5.41666666666667	0.745764394926241\\
5.83333333333333	0.771015579546663\\
6.25	0.793817274696957\\
6.66666666666667	0.814242614817971\\
7.08333333333333	0.832421019240936\\
7.5	0.848504827073016\\
7.91666666666667	0.862634546958263\\
8.33333333333333	0.874932676719704\\
8.75	0.885522987444145\\
9.16666666666667	0.894551986294129\\
9.58333333333333	0.902192104299869\\
10	0.90857659577014\\
};
\addlegendentry{$\hat{I}^\text{c}, a_0 = a_1 = 0.3$}

\addplot [color=mycolor1, dashdotted, mark=o, mark options={solid, mycolor1}, line width=0.8pt]
  table[row sep=crcr]{%
0	0.972025471259245\\
0.416666666666667	0.710429667058639\\
0.833333333333333	0.613294312999148\\
1.25	0.563068775792227\\
1.66666666666667	0.5551201575716\\
2.08333333333333	0.562434999824562\\
2.5	0.599427725591072\\
2.91666666666667	0.63063634760138\\
3.33333333333333	0.656586080413438\\
3.75	0.696994201555287\\
4.16666666666667	0.740799454485969\\
4.58333333333333	0.773443986196493\\
5	0.798151241474\\
5.41666666666667	0.823449829327684\\
5.83333333333333	0.85282558994858\\
6.25	0.883458077104579\\
6.66666666666667	0.907602568014685\\
7.08333333333333	0.924441641747133\\
7.5	0.9379114682134\\
7.91666666666667	0.949622821064049\\
8.33333333333333	0.959942899100426\\
8.75	0.969346805513485\\
9.16666666666667	0.97745927501845\\
9.58333333333333	0.983072413734669\\
10	0.985934430323197\\
};
\addlegendentry{$\hat{I}^\text{c}, a_0 = 0.3,  a_1 = 0.15$}

\end{axis}
\end{tikzpicture}%
\label{fig:FI_B}}
}
\caption{Fisher information and its estimates with: a) Gaussian input; and b) binary input.}
\label{fig:FI_Est}
\end{figure*}
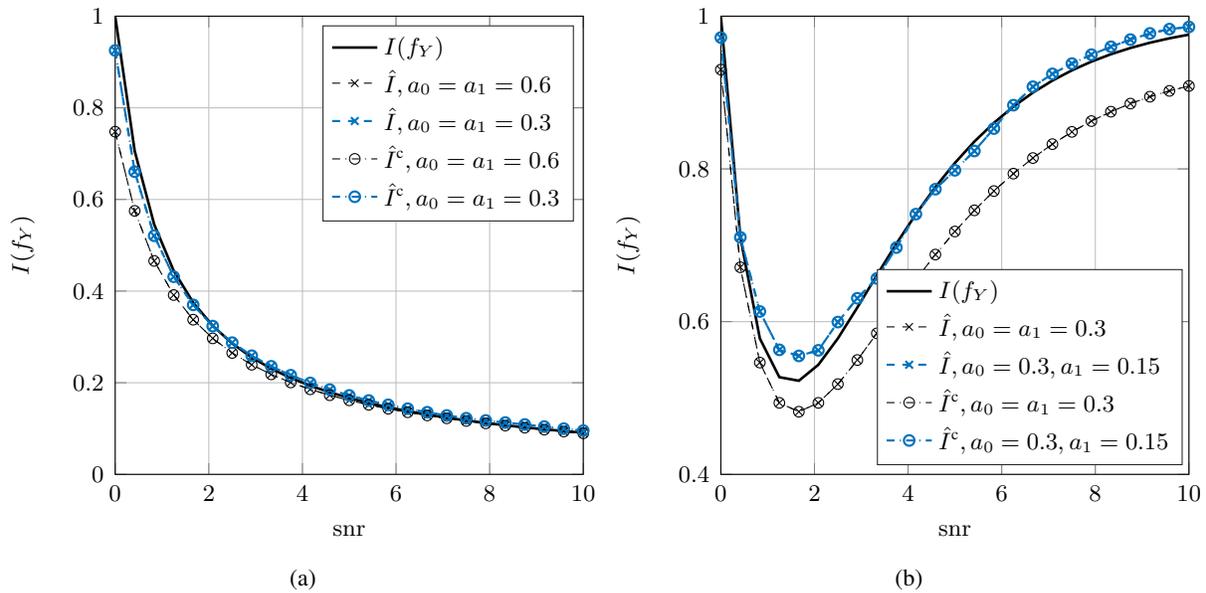

\begin{figure*}
\centering{
\subfloat[]{
%
%
\definecolor{mycolor1}{rgb}{0.00000,0.44700,0.74100}%
\begin{tikzpicture}
\pgfplotsset{every tick label/.append style={font=\footnotesize}}

\begin{axis}[%
width=2.45in,
height=2.4in,
at={(0.758in,0.481in)},
scale only axis,
xmin=0,
xmax=10,
xlabel style={font=\color{white!15!black},font=\footnotesize},
xlabel={$\snr$},
ymin=0,
ymax=1.1,
ylabel style={font=\color{white!15!black},font=\footnotesize},
ylabel={$\mmse(X,\snr)$},
axis background/.style={fill=white},
xmajorgrids,
ymajorgrids,
legend style={legend cell align=left, align=left, draw=white!15!black,font=\footnotesize}
]
\addplot [color=black, line width=1pt]
  table[row sep=crcr]{%
0	-inf\\
0.416666666666667	0.705882544656255\\
0.833333333333333	0.545453595919301\\
1.25	0.444446343425745\\
1.66666666666667	0.375000342861076\\
2.08333333333333	0.324322021854882\\
2.5	0.285712188034934\\
2.91666666666667	0.255319481336461\\
3.33333333333333	0.230771909057639\\
3.75	0.210529776598752\\
4.16666666666667	0.193550922416688\\
4.58333333333333	0.179104489219276\\
5	0.166666661605062\\
5.41666666666667	0.155844133698045\\
5.83333333333333	0.146341430972284\\
6.25	0.137926741275427\\
6.66666666666667	0.130431216565274\\
7.08333333333333	0.123709039087213\\
7.5	0.117646309969798\\
7.91666666666667	0.112150391217583\\
8.33333333333333	0.107142914783919\\
8.75	0.10256416036575\\
9.16666666666667	0.0983607033944646\\
9.58333333333333	0.0944882181811305\\
10	0.0909090958368659\\
};
\addlegendentry{$\mmse(X,\snr)$}

\addplot [color=black, dashed, mark=x, mark options={solid, black}]
  table[row sep=crcr]{%
0	inf\\
0.416666666666667	1.01978203263252\\
0.833333333333333	0.640945481922757\\
1.25	0.486849592015198\\
1.66666666666667	0.397455920376756\\
2.08333333333333	0.337519284252934\\
2.5	0.29401516414953\\
2.91666666666667	0.260800046144639\\
3.33333333333333	0.234526418864975\\
3.75	0.213175946882147\\
4.16666666666667	0.195451043372404\\
4.58333333333333	0.180489622593204\\
5	0.167680932407515\\
5.41666666666667	0.15658426564553\\
5.83333333333333	0.146875670895633\\
6.25	0.138312182389043\\
6.66666666666667	0.130706840414729\\
7.08333333333333	0.123906298368194\\
7.5	0.117786427489251\\
7.91666666666667	0.112251414901153\\
8.33333333333333	0.107224870538233\\
8.75	0.102642111794324\\
9.16666666666667	0.0984474997029717\\
9.58333333333333	0.0945936188244773\\
10	0.0910403443939829\\
};
\addlegendentry{$\hat{m}, a_0 = a_1 = 0.6$}

\addplot [color=mycolor1, dashed, mark=x, mark options={solid, mycolor1}, line width=0.8pt]
  table[row sep=crcr]{%
0	inf\\
0.416666666666667	0.814530403920237\\
0.833333333333333	0.575065593605142\\
1.25	0.454925834676938\\
1.66666666666667	0.378043186445376\\
2.08333333333333	0.32465404262693\\
2.5	0.285010103845456\\
2.91666666666667	0.254071037838986\\
3.33333333333333	0.229235379885912\\
3.75	0.208961592607971\\
4.16666666666667	0.192108189960979\\
4.58333333333333	0.177819228112374\\
5	0.165534547003079\\
5.41666666666667	0.154846217522968\\
5.83333333333333	0.145402953241289\\
6.25	0.137053586261887\\
6.66666666666667	0.129654975571846\\
7.08333333333333	0.123030031793177\\
7.5	0.116989371300545\\
7.91666666666667	0.11145376195965\\
8.33333333333333	0.106424963559397\\
8.75	0.101843587481428\\
9.16666666666667	0.0977144570787856\\
9.58333333333333	0.0939420791178543\\
10	0.0904516473020831\\
};
\addlegendentry{$\hat{m}, a_0 = a_1 = 0.3$}

\addplot [color=black, dashdotted, mark=o, mark options={solid, black}]
  table[row sep=crcr]{%
0	inf\\
0.416666666666667	1.01978203263252\\
0.833333333333333	0.640945481922757\\
1.25	0.486849592015198\\
1.66666666666667	0.397455920376756\\
2.08333333333333	0.337519284252934\\
2.5	0.29401516414953\\
2.91666666666667	0.260800046144639\\
3.33333333333333	0.234526418864975\\
3.75	0.213175946882147\\
4.16666666666667	0.195451043372404\\
4.58333333333333	0.180489622593204\\
5	0.167680932407515\\
5.41666666666667	0.15658426564553\\
5.83333333333333	0.146875670895633\\
6.25	0.138312182389043\\
6.66666666666667	0.130706840414729\\
7.08333333333333	0.123906298368194\\
7.5	0.117786427489251\\
7.91666666666667	0.112251414901153\\
8.33333333333333	0.107224870538233\\
8.75	0.102642111794324\\
9.16666666666667	0.0984474997029717\\
9.58333333333333	0.0945936188244773\\
10	0.0910403443939829\\
};
\addlegendentry{$\hat{m}^\text{c}, a_0 = a_1 = 0.6$}

\addplot [color=mycolor1, dashdotted, mark=o, mark options={solid, mycolor1}, line width=0.8pt]
  table[row sep=crcr]{%
0	inf\\
0.416666666666667	0.814530403946427\\
0.833333333333333	0.575065593605259\\
1.25	0.454925834676951\\
1.66666666666667	0.378043186445377\\
2.08333333333333	0.32465404262693\\
2.5	0.285010103845456\\
2.91666666666667	0.254071037838986\\
3.33333333333333	0.229235379885912\\
3.75	0.208961592607971\\
4.16666666666667	0.192108189960979\\
4.58333333333333	0.177819228112374\\
5	0.165534547003079\\
5.41666666666667	0.154846217522968\\
5.83333333333333	0.145402953241289\\
6.25	0.137053586261887\\
6.66666666666667	0.129654975571846\\
7.08333333333333	0.123030031793177\\
7.5	0.116989371300545\\
7.91666666666667	0.11145376195965\\
8.33333333333333	0.106424963559397\\
8.75	0.101843587481428\\
9.16666666666667	0.0977144570787856\\
9.58333333333333	0.0939420791178543\\
10	0.0904516473020831\\
};
\addlegendentry{$\hat{m}^\text{c}, a_0 = a_1 = 0.3$}

\end{axis}
\end{tikzpicture}%
}
\subfloat[]{
%
%
\definecolor{mycolor1}{rgb}{0.00000,0.44700,0.74100}%
\begin{tikzpicture}
\pgfplotsset{every tick label/.append style={font=\footnotesize}}

\begin{axis}[%
width=2.45in,
height=2.4in,
at={(0.758in,0.481in)},
scale only axis,
xmin=0,
xmax=10,
xlabel style={font=\color{white!15!black},font=\footnotesize},
xlabel={$\snr$},
ymin=0,
ymax=0.85,
ylabel style={font=\color{white!15!black},font=\footnotesize},
ylabel={$\mmse(X,\snr)$},
axis background/.style={fill=white},
xmajorgrids,
ymajorgrids,
legend style={legend cell align=left, align=left, draw=white!15!black, font=\footnotesize}
]
\addplot [color=black, line width=1pt]
  table[row sep=crcr]{%
0	-inf\\
0.416666666666667	0.694007737552566\\
0.833333333333333	0.506367110240269\\
1.25	0.378161578040566\\
1.66666666666667	0.286448895416534\\
2.08333333333333	0.219099171304888\\
2.5	0.168793026868592\\
2.91666666666667	0.130757953041349\\
3.33333333333333	0.10174184859837\\
3.75	0.0794515094002079\\
4.16666666666667	0.0622325766579848\\
4.58333333333333	0.0488468965180278\\
5	0.0384401086485242\\
5.41666666666667	0.0303112713528403\\
5.83333333333333	0.0239571092569212\\
6.25	0.0189518558448733\\
6.66666666666667	0.0150124712314327\\
7.08333333333333	0.0119063042337209\\
7.5	0.00945313930941025\\
7.91666666666667	0.00751287769331292\\
8.33333333333333	0.00597627280849083\\
8.75	0.00475790494754463\\
9.16666666666667	0.00379082403187779\\
9.58333333333333	0.00302244520017503\\
10	0.00241138969752195\\
};
\addlegendentry{$\mmse(X,\snr)$}

\addplot [color=black, dashed, mark=x, mark options={solid, black}]
  table[row sep=crcr]{%
0	inf\\
0.416666666666667	0.789199633657165\\
0.833333333333333	0.544339355654172\\
1.25	0.405069727072641\\
1.66666666666667	0.310773179848185\\
2.08333333333333	0.243058697973273\\
2.5	0.192719832603239\\
2.91666666666667	0.154374124199873\\
3.33333333333333	0.124596577317808\\
3.75	0.101214602854512\\
4.16666666666667	0.0827581217171598\\
4.58333333333333	0.0681106834377216\\
5	0.0563922510619852\\
5.41666666666667	0.0469358040136171\\
5.83333333333333	0.039254472077715\\
6.25	0.0329892360484869\\
6.66666666666667	0.0278636077773043\\
7.08333333333333	0.0236582090483385\\
7.5	0.0201993563902645\\
7.91666666666667	0.0173514256473773\\
8.33333333333333	0.0150080787936355\\
8.75	0.0130830871492406\\
9.16666666666667	0.0115034196770041\\
9.58333333333333	0.0102060412904485\\
10	0.00913768209211341\\
};
\addlegendentry{$\hat{m}, a_0 = a_1 = 0.3$}

\addplot [color=mycolor1, dashed, mark=x, mark options={solid, mycolor1}, line width=0.8pt]
  table[row sep=crcr]{%
0	inf\\
0.416666666666667	0.694968799059267\\
0.833333333333333	0.464046824401022\\
1.25	0.349544979366219\\
1.66666666666667	0.26692790545704\\
2.08333333333333	0.21003120008421\\
2.5	0.160228909763571\\
2.91666666666667	0.12663896653667\\
3.33333333333333	0.103024175875969\\
3.75	0.0808015462519234\\
4.16666666666667	0.0622081309233675\\
4.58333333333333	0.0494304030116743\\
5	0.0403697517051999\\
5.41666666666667	0.0325938776625814\\
5.83333333333333	0.0252298988659577\\
6.25	0.0186467076632673\\
6.66666666666667	0.0138596147977973\\
7.08333333333333	0.0106670623415813\\
7.5	0.00827847090487996\\
7.91666666666667	0.00636343312875168\\
8.33333333333333	0.00480685210794884\\
8.75	0.00350322222703032\\
9.16666666666667	0.00245898817980548\\
9.58333333333333	0.00176635682768673\\
10	0.00140655696768031\\
};
\addlegendentry{$\hat{m}, a_0 = 0.3,  a_1 = 0.15$}

\addplot [color=black, dashdotted, mark=o, mark options={solid, black}]
  table[row sep=crcr]{%
0	inf\\
0.416666666666667	0.789199633659584\\
0.833333333333333	0.544339355654228\\
1.25	0.405069727072654\\
1.66666666666667	0.310773179848187\\
2.08333333333333	0.243058697973274\\
2.5	0.192719832603239\\
2.91666666666667	0.154374124199873\\
3.33333333333333	0.124596577317808\\
3.75	0.101214602854512\\
4.16666666666667	0.0827581217171598\\
4.58333333333333	0.0681106834377216\\
5	0.0563922510619852\\
5.41666666666667	0.0469358040136171\\
5.83333333333333	0.039254472077715\\
6.25	0.0329892360484869\\
6.66666666666667	0.0278636077773043\\
7.08333333333333	0.0236582090483385\\
7.5	0.0201993563902645\\
7.91666666666667	0.0173514256473773\\
8.33333333333333	0.0150080787936355\\
8.75	0.0130830871492406\\
9.16666666666667	0.0115034196770041\\
9.58333333333333	0.0102060412904485\\
10	0.00914234042298603\\
};
\addlegendentry{$\hat{m}^\text{c}, a_0 = a_1 = 0.3$}

\addplot [color=mycolor1, dashdotted, mark=o, mark options={solid, mycolor1}, line width=0.8pt]
  table[row sep=crcr]{%
0	inf\\
0.416666666666667	0.694968799059267\\
0.833333333333333	0.464046824401022\\
1.25	0.349544979366219\\
1.66666666666667	0.26692790545704\\
2.08333333333333	0.21003120008421\\
2.5	0.160228909763571\\
2.91666666666667	0.12663896653667\\
3.33333333333333	0.103024175875969\\
3.75	0.0808015462519234\\
4.16666666666667	0.0622081309233675\\
4.58333333333333	0.0494304030116743\\
5	0.0403697517051999\\
5.41666666666667	0.0325938776625814\\
5.83333333333333	0.0252298988659577\\
6.25	0.0186467076632673\\
6.66666666666667	0.0138596147977973\\
7.08333333333333	0.0106670623415813\\
7.5	0.00827847090487996\\
7.91666666666667	0.00636343312875167\\
8.33333333333333	0.00480685210794884\\
8.75	0.00350322222703032\\
9.16666666666667	0.00245898817980548\\
9.58333333333333	0.00176635682768673\\
10	0.00140655696768031\\
};
\addlegendentry{$\hat{m}^\text{c}, a_0 = 0.3,  a_1 = 0.15$}

\end{axis}
\end{tikzpicture}%
}
}
\caption{MMSE and its estimates with: a) Gaussian input; and b) binary input.}
\label{fig:MMSE_Est}
\end{figure*}
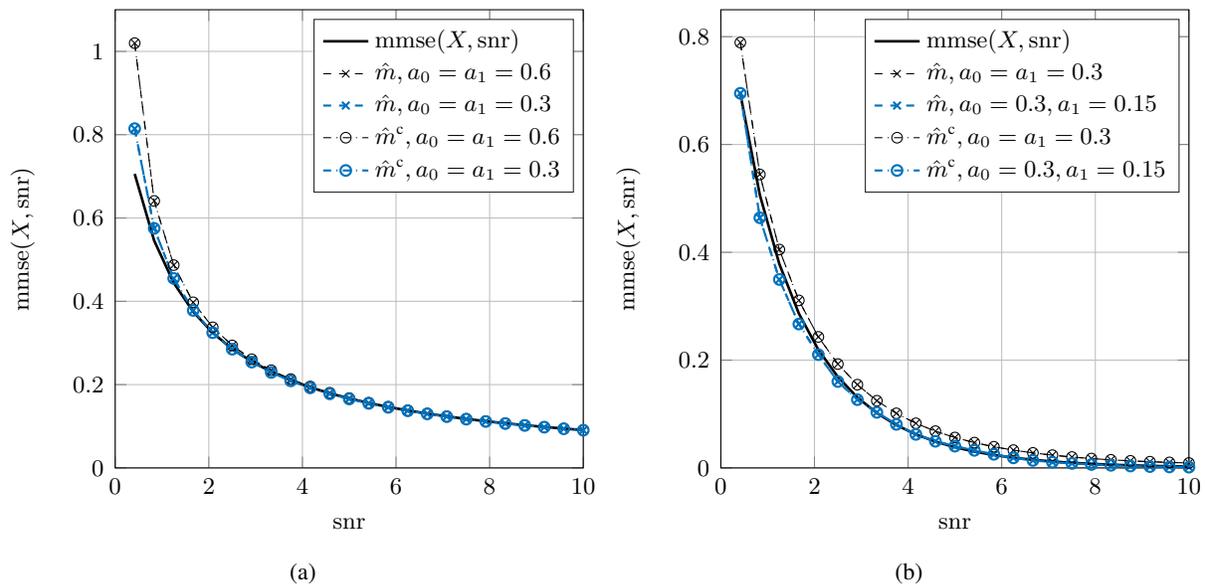

\subsection{The Estimates}\label{sec:example_est}

First, we examine the estimators $f_n$ and $f'_n$ in the two example scenarios. In both examples, we take $a = n^{-\frac{1}{8}}$ and $\snr = 1$ with $n$ varying from 50 to 5000. Note that the choices of $a_i$ and $k_n$ in Theorem~\ref{thm:FIestError_GaussianNoise} and in Theorem~\ref{thm:clipped_FIest_GaussianNoise} are not necessarily the best choices, and neither are those used in the subsequent examples. 
Figure~\ref{fig:DensityEst} shows the density function $f_Y$ and some representative realizations  $\hat{f}$ of the density estimator $f_n$ with sample size $n$ varying from 50 to 5000. As is to be expected, $f_n$ describes $f_Y$ more accurately with larger $n$.  
Similarly, Figure~\ref{fig:DerivativeEst} shows the derivative of the density function $f'_Y$ and the derivative of the density estimates  $\hat{f}'$ with sample size $n$ varying from 50 to 5000. By comparing Figure~\ref{fig:DerivativeEst} with Figure~\ref{fig:DensityEst}, we observe that $f'$ is not estimated as well as $f$. This is a natural consequence of plug-in methods.

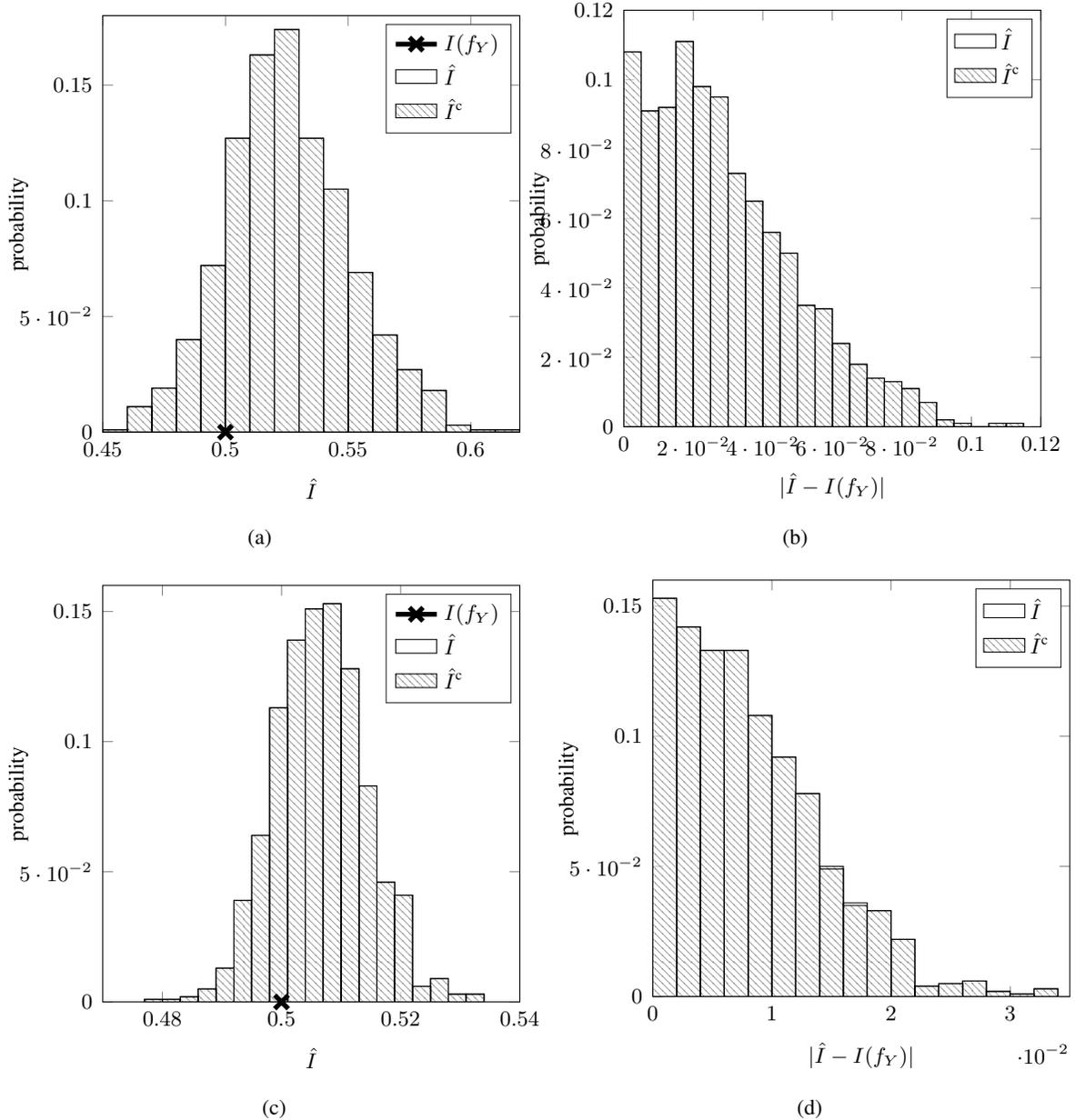
\begin{figure*}[!t]
\centering{
\subfloat[]{
%
%
\definecolor{mycolor1}{rgb}{0.00000,0.44700,0.74100}%
\begin{tikzpicture}
\pgfplotsset{every tick label/.append style={font=\footnotesize}}

\begin{axis}[%
width=2.4in,
height=2.4in,
at={(0.758in,0.504in)},
scale only axis,
xmin=0.45,
xmax=0.62,
xlabel style={font=\color{white!15!black},font=\footnotesize},
xlabel={$\hat{I}$},
ymin=0,
ymax=0.18,
ylabel style={font=\color{white!15!black},font=\footnotesize},
ylabel={probability},
axis background/.style={fill=white},
legend style={legend cell align=left, align=left, draw=white!15!black,font=\footnotesize}
]
\addplot [color=black, line width=2.0pt, draw=none, mark size=4.0pt, mark=x, mark options={solid, black}]
  table[row sep=crcr]{%
0.499999564370385	0\\
};
\addlegendentry{$I(f_Y)$}

\addplot[ybar interval, fill=white, fill opacity=0.5, draw=black, area legend] table[row sep=crcr] {%
x	y\\
0.45	0.001\\
0.46	0.011\\
0.47	0.019\\
0.48	0.04\\
0.49	0.072\\
0.5	0.127\\
0.51	0.163\\
0.52	0.174\\
0.53	0.127\\
0.54	0.105\\
0.55	0.069\\
0.56	0.042\\
0.57	0.027\\
0.58	0.018\\
0.59	0.003\\
0.6	0.001\\
0.61	0.001\\
0.62	0.001\\
};
\addlegendentry{$\hat{I}$}

\addplot[ybar interval, pattern=north west lines, pattern color=black, fill opacity=0.6, draw=black, area legend] table[row sep=crcr] {%
x	y\\
0.45	0.001\\
0.46	0.011\\
0.47	0.019\\
0.48	0.04\\
0.49	0.072\\
0.5	0.127\\
0.51	0.163\\
0.52	0.174\\
0.53	0.127\\
0.54	0.105\\
0.55	0.069\\
0.56	0.042\\
0.57	0.027\\
0.58	0.018\\
0.59	0.003\\
0.6	0.001\\
0.61	0.001\\
0.62	0.001\\
};
\addlegendentry{$\hat{I}^\text{c}$}

\end{axis}
\end{tikzpicture}%
}
\subfloat[]{
%
%
\definecolor{mycolor1}{rgb}{0.00000,0.44700,0.74100}%
\begin{tikzpicture}
\pgfplotsset{every tick label/.append style={font=\footnotesize}}

\begin{axis}[%
width=2.4in,
height=2.4in,
at={(0.758in,0.481in)},
scale only axis,
xmin=0,
xmax=0.12,
xlabel style={font=\color{white!15!black},font=\footnotesize},
xlabel={$|\hat{I} - I(f_Y)|$},
ymin=0,
ymax=0.12,
ylabel style={font=\color{white!15!black},font=\footnotesize},
ylabel={probability},
axis background/.style={fill=white},
legend style={legend cell align=left, align=left, draw=white!15!black,font=\footnotesize}
]
\addplot[ybar interval, fill=white, fill opacity=0.5, draw=black, area legend] table[row sep=crcr] {%
x	y\\
0	0.108\\
0.005	0.091\\
0.01	0.092\\
0.015	0.111\\
0.02	0.098\\
0.025	0.095\\
0.03	0.073\\
0.035	0.065\\
0.04	0.056\\
0.045	0.05\\
0.05	0.035\\
0.055	0.034\\
0.06	0.024\\
0.065	0.018\\
0.07	0.014\\
0.075	0.013\\
0.08	0.011\\
0.085	0.007\\
0.09	0.002\\
0.095	0.001\\
0.1	0\\
0.105	0.001\\
0.11	0.001\\
0.115	0.001\\
};
\addlegendentry{$\hat{I}$}

\addplot[ybar interval, pattern=north west lines, pattern color=black, fill opacity=0.6, draw=black, area legend] table[row sep=crcr] {%
x	y\\
0	0.108\\
0.005	0.091\\
0.01	0.092\\
0.015	0.111\\
0.02	0.098\\
0.025	0.095\\
0.03	0.073\\
0.035	0.065\\
0.04	0.056\\
0.045	0.05\\
0.05	0.035\\
0.055	0.034\\
0.06	0.024\\
0.065	0.018\\
0.07	0.014\\
0.075	0.013\\
0.08	0.011\\
0.085	0.007\\
0.09	0.002\\
0.095	0.001\\
0.1	0\\
0.105	0.001\\
0.11	0.001\\
0.115	0.001\\
};
\addlegendentry{$\hat{I}^\text{c}$}

\end{axis}
\end{tikzpicture}%
}\\
\subfloat[]{
%
%
\definecolor{mycolor1}{rgb}{0.00000,0.44700,0.74100}%
\begin{tikzpicture}
\pgfplotsset{every tick label/.append style={font=\footnotesize}}

\begin{axis}[%
width=2.4in,
height=2.4in,
at={(0.758in,0.504in)},
scale only axis,
xmin=0.47,
xmax=0.54,
xlabel style={font=\color{white!15!black},font=\footnotesize},
xlabel={$\hat{I}$},
ymin=0,
ymax=0.16,
ylabel style={font=\color{white!15!black},font=\footnotesize},
ylabel={probability},
axis background/.style={fill=white},
legend style={legend cell align=left, align=left, draw=white!15!black,font=\footnotesize}
]
\addplot [color=black, line width=2.0pt, draw=none, mark size=4.0pt, mark=x, mark options={solid, black}]
  table[row sep=crcr]{%
0.499999564370385	0\\
};
\addlegendentry{$I(f_Y)$}

\addplot[ybar interval, fill=white, fill opacity=0.5, draw=black, area legend] table[row sep=crcr] {%
x	y\\
0.477	0.001\\
0.48	0.001\\
0.483	0.002\\
0.486	0.005\\
0.489	0.013\\
0.492	0.039\\
0.495	0.064\\
0.498	0.113\\
0.501	0.139\\
0.504	0.151\\
0.507	0.153\\
0.51	0.128\\
0.513	0.083\\
0.516	0.046\\
0.519	0.041\\
0.522	0.006\\
0.525	0.009\\
0.528	0.003\\
0.531	0.003\\
0.534	0.003\\
};
\addlegendentry{$\hat{I}$}

\addplot[ybar interval, pattern=north west lines, pattern color=black, fill opacity=0.6, draw=black, area legend] table[row sep=crcr] {%
x	y\\
0.477	0.001\\
0.48	0.001\\
0.483	0.002\\
0.486	0.005\\
0.489	0.013\\
0.492	0.039\\
0.495	0.064\\
0.498	0.113\\
0.501	0.139\\
0.504	0.151\\
0.507	0.153\\
0.51	0.128\\
0.513	0.083\\
0.516	0.046\\
0.519	0.041\\
0.522	0.006\\
0.525	0.009\\
0.528	0.003\\
0.531	0.003\\
0.534	0.003\\
};
\addlegendentry{$\hat{I}^\text{c}$}

\end{axis}
\end{tikzpicture}%
}
\subfloat[]{
%
%
\definecolor{mycolor1}{rgb}{0.00000,0.44700,0.74100}%
\begin{tikzpicture}
\pgfplotsset{every tick label/.append style={font=\footnotesize}}

\begin{axis}[%
width=2.4in,
height=2.4in,
at={(0.758in,0.481in)},
scale only axis,
xmin=0,
xmax=0.035,
xlabel style={font=\color{white!15!black},font=\footnotesize},
xlabel={$|\hat{I} - I(f_Y)|$},
ymin=0,
ymax=0.16,
ylabel style={font=\color{white!15!black},font=\footnotesize},
ylabel={probability},
axis background/.style={fill=white},
legend style={legend cell align=left, align=left, draw=white!15!black,font=\footnotesize}
]
\addplot[ybar interval, fill=white, fill opacity=0.5, draw=black, area legend] table[row sep=crcr] {%
x	y\\
0	0.153\\
0.002	0.142\\
0.004	0.133\\
0.006	0.133\\
0.008	0.108\\
0.01	0.092\\
0.012	0.078\\
0.014	0.049\\
0.016	0.036\\
0.018	0.033\\
0.02	0.022\\
0.022	0.004\\
0.024	0.005\\
0.026	0.006\\
0.028	0.002\\
0.03	0.001\\
0.032	0.003\\
0.034	0.003\\
};
\addlegendentry{$\hat{I}$}

\addplot[ybar interval, pattern=north west lines, pattern color=black, fill opacity=0.6, draw=black, area legend] table[row sep=crcr] {%
x	y\\
0	0.153\\
0.002	0.142\\
0.004	0.133\\
0.006	0.133\\
0.008	0.108\\
0.01	0.092\\
0.012	0.078\\
0.014	0.05\\
0.016	0.035\\
0.018	0.033\\
0.02	0.022\\
0.022	0.004\\
0.024	0.005\\
0.026	0.006\\
0.028	0.002\\
0.03	0.001\\
0.032	0.003\\
0.034	0.003\\
};
\addlegendentry{$\hat{I}^\text{c}$}

\end{axis}
\end{tikzpicture}%
}
}
\caption{Comparison of the Fisher information and the  estimates with Gaussian input: a) histograms of the estimates with $n=10^3$; b) histograms of errors of the  estimates with $n=10^3$; c) histograms of the  estimates with $n=10^4$; and d) histograms of errors of the estimates with $n=10^4$.}
\label{fig:sim_G}
\end{figure*}

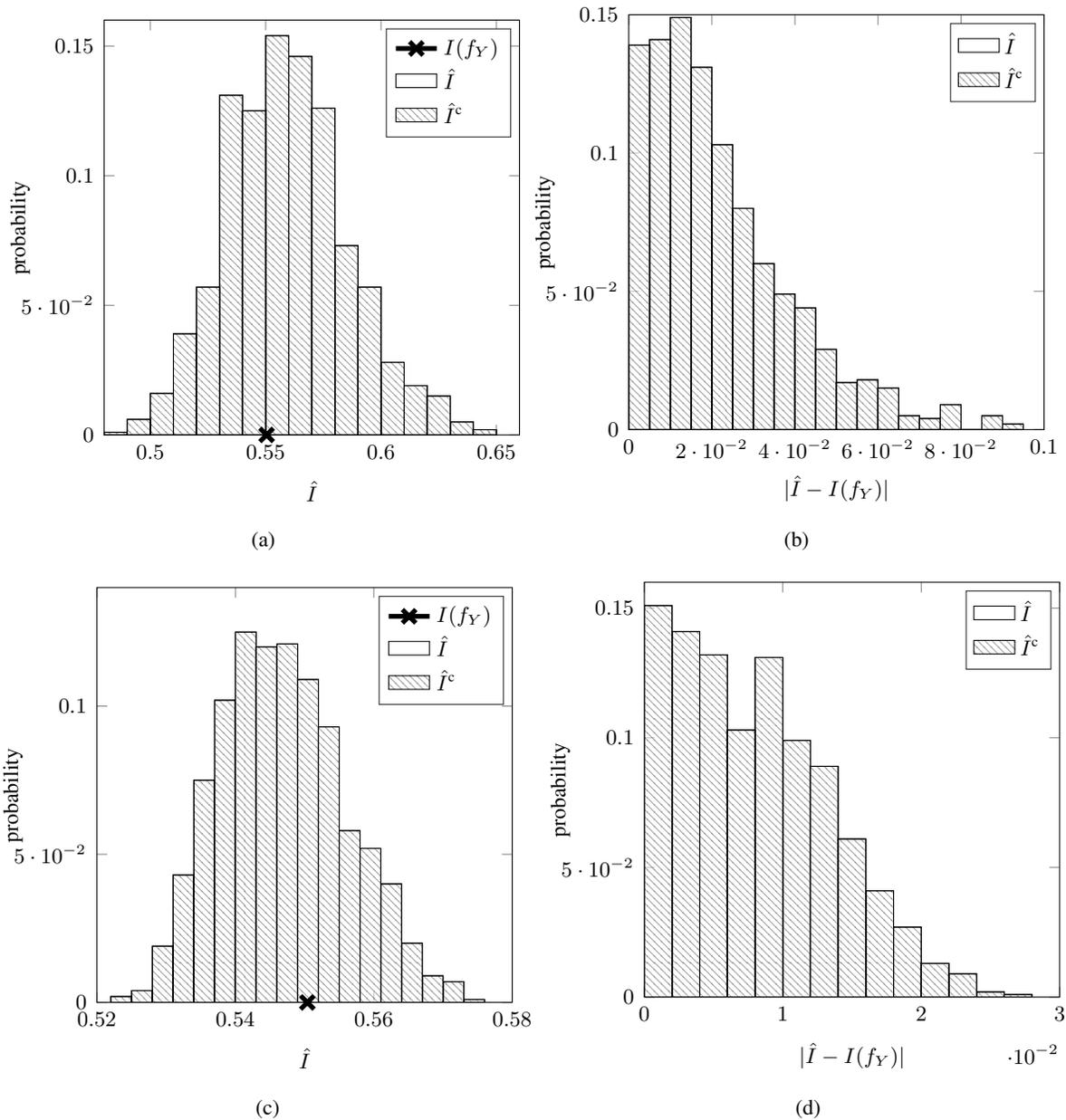
\begin{figure*}[!t]
\centering{
\subfloat[]{
%
%
\definecolor{mycolor1}{rgb}{0.00000,0.44700,0.74100}%
\begin{tikzpicture}
\pgfplotsset{every tick label/.append style={font=\footnotesize}}

\begin{axis}[%
width=2.4in,
height=2.4in,
at={(0.758in,0.504in)},
scale only axis,
xmin=0.48,
xmax=0.66,
xlabel style={font=\color{white!15!black},font=\footnotesize},
xlabel={$\hat{I}$},
ymin=0,
ymax=0.16,
ylabel style={font=\color{white!15!black},font=\footnotesize},
ylabel={probability},
axis background/.style={fill=white},
legend style={legend cell align=left, align=left, draw=white!15!black,font=\footnotesize}
]
\addplot [color=black, line width=2.0pt, draw=none, mark size=4.0pt, mark=x, mark options={solid, black}]
  table[row sep=crcr]{%
0.55039711331101	0\\
};
\addlegendentry{$I(f_Y)$}

\addplot[ybar interval, fill=white, fill opacity=0.5, draw=black, area legend] table[row sep=crcr] {%
x	y\\
0.48	0.001\\
0.49	0.006\\
0.5	0.016\\
0.51	0.039\\
0.52	0.057\\
0.53	0.131\\
0.54	0.125\\
0.55	0.154\\
0.56	0.146\\
0.57	0.126\\
0.58	0.073\\
0.59	0.057\\
0.6	0.028\\
0.61	0.019\\
0.62	0.015\\
0.63	0.005\\
0.64	0.002\\
0.65	0.002\\
};
\addlegendentry{$\hat{I}$}

\addplot[ybar interval, pattern=north west lines, pattern color=black, fill opacity=0.6, draw=black, area legend] table[row sep=crcr] {%
x	y\\
0.48	0.001\\
0.49	0.006\\
0.5	0.016\\
0.51	0.039\\
0.52	0.057\\
0.53	0.131\\
0.54	0.125\\
0.55	0.154\\
0.56	0.146\\
0.57	0.126\\
0.58	0.073\\
0.59	0.057\\
0.6	0.028\\
0.61	0.019\\
0.62	0.015\\
0.63	0.005\\
0.64	0.002\\
0.65	0.002\\
};
\addlegendentry{$\hat{I}^\text{c}$}

\end{axis}
\end{tikzpicture}%
}
\subfloat[]{
%
%
\definecolor{mycolor1}{rgb}{0.00000,0.44700,0.74100}%
\begin{tikzpicture}
\pgfplotsset{every tick label/.append style={font=\footnotesize}}

\begin{axis}[%
width=2.4in,
height=2.4in,
at={(0.758in,0.481in)},
scale only axis,
xmin=0,
xmax=0.1,
xlabel style={font=\color{white!15!black},font=\footnotesize},
xlabel={$|\hat{I} - I(f_Y)|$},
ymin=0,
ymax=0.15,
ylabel style={font=\color{white!15!black},font=\footnotesize},
ylabel={probability},
axis background/.style={fill=white},
legend style={legend cell align=left, align=left, draw=white!15!black,font=\footnotesize}
]
\addplot[ybar interval, fill=white, fill opacity=0.5, draw=black, area legend] table[row sep=crcr] {%
x	y\\
0	0.139\\
0.005	0.141\\
0.01	0.149\\
0.015	0.131\\
0.02	0.103\\
0.025	0.08\\
0.03	0.06\\
0.035	0.049\\
0.04	0.044\\
0.045	0.029\\
0.05	0.017\\
0.055	0.018\\
0.06	0.015\\
0.065	0.005\\
0.07	0.004\\
0.075	0.009\\
0.08	0\\
0.085	0.005\\
0.09	0.002\\
0.095	0.002\\
};
\addlegendentry{$\hat{I}$}

\addplot[ybar interval, pattern=north west lines, pattern color=black, fill opacity=0.6, draw=black, area legend] table[row sep=crcr] {%
x	y\\
0	0.139\\
0.005	0.141\\
0.01	0.149\\
0.015	0.131\\
0.02	0.103\\
0.025	0.08\\
0.03	0.06\\
0.035	0.049\\
0.04	0.044\\
0.045	0.029\\
0.05	0.017\\
0.055	0.018\\
0.06	0.015\\
0.065	0.005\\
0.07	0.004\\
0.075	0.009\\
0.08	0\\
0.085	0.005\\
0.09	0.002\\
0.095	0.002\\
};
\addlegendentry{$\hat{I}^\text{c}$}

\end{axis}
\end{tikzpicture}%
}\\
\subfloat[]{
%
%
\definecolor{mycolor1}{rgb}{0.00000,0.44700,0.74100}%
\begin{tikzpicture}
\pgfplotsset{every tick label/.append style={font=\footnotesize}}

\begin{axis}[%
width=2.4in,
height=2.4in,
at={(0.758in,0.504in)},
scale only axis,
xmin=0.52,
xmax=0.58,
xlabel style={font=\color{white!15!black},font=\footnotesize},
xlabel={$\hat{I}$},
ymin=0,
ymax=0.14,
ylabel style={font=\color{white!15!black},font=\footnotesize},
ylabel={probability},
axis background/.style={fill=white},
legend style={legend cell align=left, align=left, draw=white!15!black,font=\footnotesize}
]
\addplot [color=black, line width=2.0pt, draw=none, mark size=4.0pt, mark=x, mark options={solid, black}]
  table[row sep=crcr]{%
0.55039711331101	0\\
};
\addlegendentry{$I(f_Y)$}

\addplot[ybar interval, fill=white, fill opacity=0.5, draw=black, area legend] table[row sep=crcr] {%
x	y\\
0.522	0.002\\
0.525	0.004\\
0.528	0.019\\
0.531	0.043\\
0.534	0.075\\
0.537	0.102\\
0.54	0.125\\
0.543	0.12\\
0.546	0.121\\
0.549	0.109\\
0.552	0.093\\
0.555	0.058\\
0.558	0.052\\
0.561	0.04\\
0.564	0.02\\
0.567	0.009\\
0.57	0.007\\
0.573	0.001\\
0.576	0.001\\
};
\addlegendentry{$\hat{I}$}

\addplot[ybar interval, pattern=north west lines, pattern color=black, fill opacity=0.6, draw=black, area legend] table[row sep=crcr] {%
x	y\\
0.522	0.002\\
0.525	0.004\\
0.528	0.019\\
0.531	0.043\\
0.534	0.075\\
0.537	0.102\\
0.54	0.125\\
0.543	0.12\\
0.546	0.121\\
0.549	0.109\\
0.552	0.093\\
0.555	0.058\\
0.558	0.052\\
0.561	0.04\\
0.564	0.02\\
0.567	0.009\\
0.57	0.007\\
0.573	0.001\\
0.576	0.001\\
};
\addlegendentry{$\hat{I}^\text{c}$}

\end{axis}
\end{tikzpicture}%
}
\subfloat[]{
%
%
\definecolor{mycolor1}{rgb}{0.00000,0.44700,0.74100}%
\begin{tikzpicture}
\pgfplotsset{every tick label/.append style={font=\footnotesize}}

\begin{axis}[%
width=2.4in,
height=2.4in,
at={(0.758in,0.481in)},
scale only axis,
xmin=0,
xmax=0.03,
xlabel style={font=\color{white!15!black},font=\footnotesize},
xlabel={$|\hat{I} - I(f_Y)|$},
ymin=0,
ymax=0.16,
ylabel style={font=\color{white!15!black},font=\footnotesize},
ylabel={probability},
axis background/.style={fill=white},
legend style={legend cell align=left, align=left, draw=white!15!black,font=\footnotesize}
]
\addplot[ybar interval, fill=white, fill opacity=0.5, draw=black, area legend] table[row sep=crcr] {%
x	y\\
0	0.151\\
0.002	0.141\\
0.004	0.132\\
0.006	0.103\\
0.008	0.131\\
0.01	0.099\\
0.012	0.089\\
0.014	0.061\\
0.016	0.041\\
0.018	0.027\\
0.02	0.013\\
0.022	0.009\\
0.024	0.002\\
0.026	0.001\\
0.028	0.001\\
};
\addlegendentry{$\hat{I}$}

\addplot[ybar interval, pattern=north west lines, pattern color=black, fill opacity=0.6, draw=black, area legend] table[row sep=crcr] {%
x	y\\
0	0.151\\
0.002	0.141\\
0.004	0.132\\
0.006	0.103\\
0.008	0.131\\
0.01	0.099\\
0.012	0.089\\
0.014	0.061\\
0.016	0.041\\
0.018	0.027\\
0.02	0.013\\
0.022	0.009\\
0.024	0.002\\
0.026	0.001\\
0.028	0.001\\
};
\addlegendentry{$\hat{I}^\text{c}$}

\end{axis}
\end{tikzpicture}%
}
}
\caption{Comparison of the Fisher information and the estimates with binary input: a) histograms of the v with $n=10^3$; b) histograms of errors of the  estimates with $n=10^3$; c) histograms of the estimates with $n=10^4$; and d) histograms of errors of the estimates with $n=10^4$.}
\label{fig:sim_B}
\end{figure*}

Second, let us examine the Fisher information estimators $I_n$ and $I_n^{\text{c}}$. Here, we take $n = 10^4$ and $k_n = 10$. Figure~\ref{fig:FI_Est} shows the Fisher information and the estimates for different values of $a_0, a_1$ when $\snr$ varies from 1 to 10. From the results, we can see that $I_n$ coincides with $I_n^\text{c}$, since $\rho_n$ rarely exceeds $\rho_{\max}$ in these examples. As a result, we can say that the better bounds on the convergence rate of the clipped estimator do not necessarily come at the expense of a sacrifice in accuracy compared to the Bhattacharya estimator. 
Moreover, Figure~\ref{fig:FI_B} demonstrates that small bandwidths can lead to under-smoothing (over-estimation of the Fisher information) while large bandwidths can lead to over-smoothing (under-estimation of the Fisher information).

Next, we examine the MMSE estimators $\mmse_n$ and $\mmse_n^\text{c}$  and denote their corresponding estimates by $\hat{m}$ and $\hat{m}^c$ respectively. Again, we take $n = 10^4$ and $k_n = 10$. Figure~\ref{fig:MMSE_Est} shows the MMSE and its estimates with different values of $a_0, a_1$ when $\snr$ varies from 1 to 10. The observations are similar to those of Figure~\ref{fig:FI_Est}.

\subsection{Bias and Variance}\label{sec:example_bias}

To take a closer look at the performance of the Fisher information estimators, we next present some additional simulation results. Generally, we repeat the simulation experiments for $T = 10^3$ times and then plot the corresponding histograms of the estimates as well as the histograms of the errors. In addition, we set $a=n^{1/6}$, and $k_n=\log(n)$. Figure~\ref{fig:sim_G} and Figure~\ref{fig:sim_B} show the histograms for Gaussian input and binary input, respectively. Again, in both figures, there is no obvious difference between the Bhattacharya estimator $I_n$ and the clipped estimator $I^\text{c}_n$ due to the fact that $\rho_{\max}$ dominates $\rho_n$ in these examples. Moreover, in both examples, the errors are reduced more than 50\% when $n$ increases from $10^3$ to $10^4$.

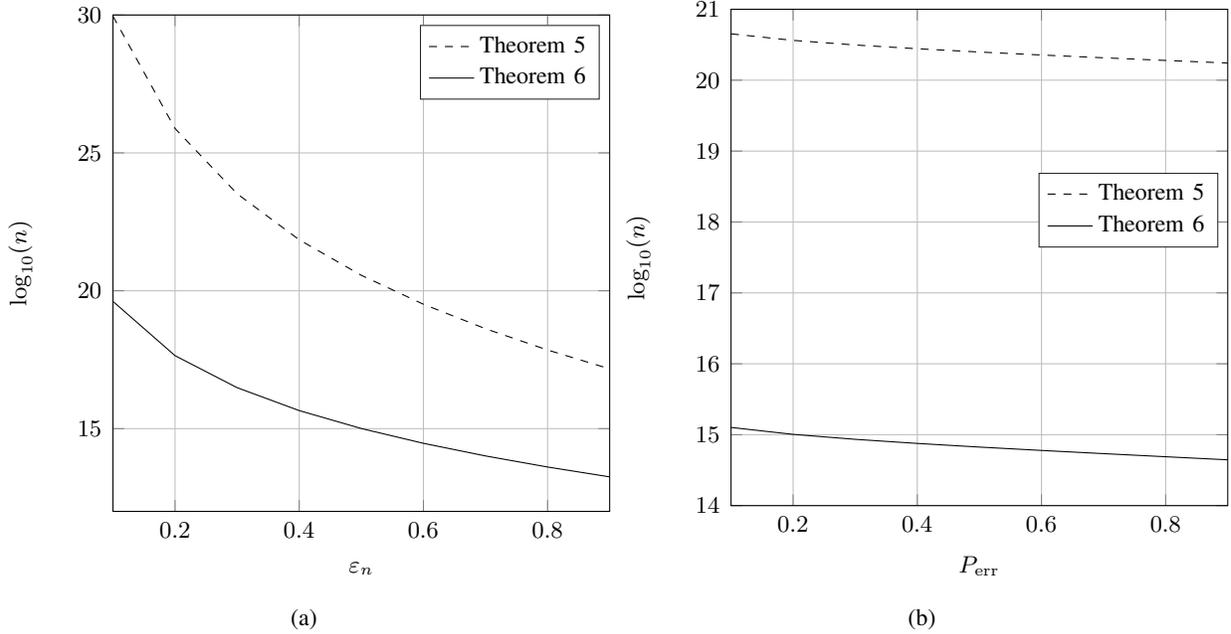
\begin{figure*}
\centering{
\subfloat[]{
%
%
\begin{tikzpicture}
\pgfplotsset{every tick label/.append style={font=\footnotesize}}

\begin{axis}[%
width=2.6in,
height=2.6in,
at={(0.758in,0.481in)},
scale only axis,
xmin=0.1,
xmax=0.9,
xlabel style={font=\color{white!15!black},font=\footnotesize},
xlabel={$\varepsilon_n$},
ymin=12,
ymax=30,
ylabel style={font=\color{white!15!black},font=\footnotesize},
ylabel={$\log_{10}(n)$},
axis background/.style={fill=white},
xmajorgrids,
ymajorgrids,
legend style={legend cell align=left, align=left, draw=white!15!black,font=\footnotesize}
]

\addplot [color=black, dashed]
  table[row sep=crcr]{%
0.1	29.957378059561\\
0.2	25.8841098674962\\
0.3	23.5173799893971\\
0.4	21.8491904122939\\
0.5	20.5612603866351\\
0.6	19.511353765118\\
0.7	18.6234938991055\\
0.8	17.8522827104868\\
0.9	17.1683707929747\\
};
\addlegendentry{Theorem~\ref{thm:FIestError_GaussianNoise}}

\addplot [color=black]
  table[row sep=crcr]{%
0.1	19.6148639373318\\
0.2	17.646724713343\\
0.3	16.4857418842249\\
0.4	15.6549356215473\\
0.5	15.0047063054141\\
0.6	14.4683606678186\\
0.7	14.0102748686814\\
0.8	13.6091374292787\\
0.9	13.2511515414078\\
};
\addlegendentry{Theorem~\ref{thm:clipped_FIest_GaussianNoise}}

\end{axis}
\end{tikzpicture}%
\label{fig:n_err}}
\hfil
\subfloat[]{
%
%
\begin{tikzpicture}
\pgfplotsset{every tick label/.append style={font=\footnotesize}}

\begin{axis}[%
width=2.6in,
height=2.6in,
at={(0.758in,0.481in)},
scale only axis,
xmin=0.1,
xmax=0.9,
xlabel style={font=\color{white!15!black},font=\footnotesize},
xlabel={$P_{\rm err}$},
ymin=14,
ymax=21,
ylabel style={font=\color{white!15!black},font=\footnotesize},
ylabel={$\log_{10}(n)$},
axis background/.style={fill=white},
xmajorgrids,
ymajorgrids,
legend style={at={(0.8,0.52)}, anchor=south, legend cell align=left, align=left, draw=white!15!black,font=\footnotesize}
]
\addplot [color=black, dashed]
  table[row sep=crcr]{%
0.1	20.6536133932348\\
0.2	20.5612603866351\\
0.3	20.4962693145743\\
0.4	20.4432618658417\\
0.5	20.3969810189517\\
0.6	20.3549137689003\\
0.7	20.3156173892267\\
0.8	20.2781490879136\\
0.9	20.24182079047\\
};
\addlegendentry{Theorem~\ref{thm:FIestError_GaussianNoise}}

\addplot [color=black]
  table[row sep=crcr]{%
0.1	15.1024603139813\\
0.2	15.0047063054141\\
0.3	14.9348159899415\\
0.4	14.8769355617011\\
0.5	14.8256024552098\\
0.6	14.7781671316714\\
0.7	14.7330695708546\\
0.8	14.689249008401\\
0.9	14.6458868371555\\
};
\addlegendentry{Theorem~\ref{thm:clipped_FIest_GaussianNoise}}

\end{axis}
\end{tikzpicture}%
\label{fig:n_Perr}}
}
\caption{Sample complexity with Gaussian input: a) number of samples required versus error of the estimators $ I_n$ and $ I_n^\text{c}$ given $P_{\rm err}=0.2$; and b) number of samples required versus confidence of the estimators with given $\varepsilon_n = 0.5$.}
\label{fig:complexity}
\end{figure*}

\subsection{Sample Complexity}

Finally, we would like to demonstrate the difference in the bounds on the convergence rates between Bhattacharya's estimator and its clipped version by showing sample complexity of the two estimators, that is, the required number of samples to guarantee a given accuracy with a given confidence. In order to make the compariosn as fair as possible, the estimator parameters, including $a_i$ and $k_n$, are not chosen according to Theorem~\ref{thm:FIestError_GaussianNoise} or Theorem~\ref{thm:clipped_FIest_GaussianNoise}. Instead, we numerically compute the optimal parameters for each case. Let $P_{\rm err} = \mathbb{P} \left[ \left| I_n - I(f_Y) \right| \ge \varepsilon_n \right]$. Figure~\ref{fig:n_err} shows the corresponding bounds on the sample complexities of the two estimator with $P_{\rm err}=0.2$ and $\varepsilon_n$ varying from 0.1 to 0.9. Note that the results with larger $\varepsilon_n$ are not shown since $I(f_Y) \le 1$ as shown in Lemma~\ref{lem:FIestErrGaussianNoise}. Moreover, Figure~\ref{fig:n_Perr} shows the sample complexities for $\varepsilon_n = 0.5$ with $P_{\rm err}$ varying from 0.1 to 0.9. By inspection, it is clear that the clipped estimator significantly reduces the sample complexity without sacrifices in performance (as shown in Section~\ref{sec:example_est} and Section~\ref{sec:example_bias}). 
The comparisons of complexities in the binary example are omitted since the results are similar to those of the Gaussian example.

\section{Conclusion}
\label{sec:conclusion}

This work has focused on the estimation of the Fisher information for location of a random variable based on plug-in estimators of the density and its derivative. The paper has considered two estimators of the Fisher information. The first estimator is the estimator due to  Bhattacharya. For this estimator, new sharper convergence results have been provided. 
The paper has also proposed a second estimator, termed clipped estimator, which provides better bounds on the convergence rates than the Bhattacharya estimator. 
The results of both estimators have been specialized to the practically relevant case of a Gaussian noise contaminated random variable.  Moreover,  using special proprieties of the Gaussian noise case, an estimator for the minimum mean square error (MMSE) has been proposed, and the convergence rates have been analyzed. This was done by using Brown's identity, which connects the Fisher information and the MMSE. 
 

\begin{appendices}

\section{A Proof of Theorem~\ref{thm:boundsEstDensityandDerivatives} } 
\label{app:thm_boundsEstDensityandDerivatives}

Our starting point is the following bound due to \cite[p.1188]{schuster1969estimation}:
\begin{align}
    \sup_{t \in \mathbb{R}} \left| \mathbb{E}\left[f_n^{(r)}(t)\right]- f_n^{(r)}(t) \right| 
    &\le \frac{ v_r}{a^{r+1}} \sup_{t \in \mathbb{R}} \left| F_n(t)-F_Y(t) \right|,
    \label{eq:cdfBound}
\end{align} 
where  $F$ is the CDF  of $f$, $F_n$ is the empirical CDF, and $v_r$ is defined in \eqref{eq:def_vr}. Now let $\delta_{r,a}$ be as in \eqref{eq:def_deltara}, and consider the following sequence of bounds: 
\begin{align}
    \mathbb{P} \left[ \sup_{t \in \mathbb{R}} \left| f_n^{(r)}(t)-f^{(r)}(t) \right| >\epsilon  \right]  
    &\le \mathbb{P} \left[ \sup_{t \in \mathbb{R}} \left| f_n^{(r)}(t)-\E[f_n^{(r)}(t)] \right| >\epsilon-\delta_{r,a} \right] 
    \label{eq:TriangleInequaltiy}\\
    &\le \mathbb{P} \left[ \sup_{t \in \mathbb{R}} | F_n(t)-F(t) | > \frac{a^{r+1} (\epsilon- \delta_{r,a})}{ v_r}  \right] 
    \label{eq:Applyingcdfbound}\\
    &\le 2 {\rm e}^{-2n \frac{a^{2r+2} (\epsilon- \delta_{r,a})^2}{ v_r^2} }, \label{eq:BoundsOnProbabilityOfDeviation}
\end{align}
where \eqref{eq:TriangleInequaltiy}  follows by using the triangle inequality;  \eqref{eq:Applyingcdfbound} follows by using the bound in \eqref{eq:cdfBound}; and \eqref{eq:BoundsOnProbabilityOfDeviation} follows by using the sharp DKW inequality  \cite{massart1990tight}
\begin{align}
    \mathbb{P} \left[ \sup_{t \in \mathbb{R}} | F_n(t)-F(t) | > \epsilon \right] 
    \le 2 {\rm e}^{-2 n \epsilon^2}.
\end{align}
This concludes the proof.

\section{A Proof of Theorem~\ref{thm:FIestError}}
\label{app:thm_FIestError}

First,  using the triangle inequality we have that 
\begin{align}
    \left |I(f) -  I_{n} \right|  
    \le  \left| \int_{|t| \le k_n} \frac{(f_n'(t))^2}{f_n(t)}- \frac{(f'(t))^2}{f(t)} \dr t  \right|   +  c(k_n).  
    \label{eq:DiffOfFishersManipulation}
\end{align}
Next, we bound the first term in \eqref{eq:DiffOfFishersManipulation} 
\begin{align}
    & \left| \int_{|t| \le k_n} \frac{(f_n'(t))^2}{f_n(t)}- \frac{(f'(t))^2}{f(t)} \dr t  \right| \notag\\
    &= \left| \int_{|t| \le k_n} \frac{f(t) (f_n'(t))^2- f_n(t) (f'(t))^2}{ f_n(t) f(t)}  \dr t  \right| \\
    &\leq  \left| \int_{|t| \le k_n} \frac{f(t) (f_n'(t))^2- f(t) (f'(t))^2}{ f_n(t) f(t)}  \dr t  \right|  + \left| \int_{|t| \le k_n} \frac{f(t) (f'(t))^2- f_n(t) (f'(t))^2}{f_n(t) f(t)}  \dr t  \right| \\
    &= \left| \int_{|t| \le k_n} \frac{ (f_n'(t))^2- (f'(t))^2 }{ f_n(t)}  \dr t  \right|+ \left| \int_{|t| \le k_n} \frac{ f_n(t)-f(t)}{ f_n(t)} \frac{(f'(t))^2}{f(t)}  \dr t  \right|\\
    &\le \sup_{|t| \le k_n} \frac{|f_n'(t)+ f'(t)| }{f_n(t)} \left| f_n(t)-f(t)\right|  2 k_n  + \sup_{|t| \le k_n} \frac{ |f_n(t)-f(t)| }{ f_n(t)} \int_{|t| \le k_n} \frac{ (f'(t))^2}{  f(t)} \dr t \\
    & \le \sup_{|t| \le k_n} \frac{|f_n'(t)+ f'(t)| }{f_n(t)} \epsilon_1 2 k_n + \sup_{|t| \le k_n} \frac{ 1 }{ f_n(t)} \epsilon_0 I(f), 
    \label{eq:ClippedDifferenceBound}
\end{align}
where the last bound follows from the assumptions in \eqref{eq:epsilon_i}. 
Now consider the first term in \eqref{eq:ClippedDifferenceBound} 
\begin{align}
    \sup_{|t| \le k_n} \frac{|f_n'(t)+ f'(t)| }{f_n(t)} 
    &\le \sup_{|t| \le k_n} \frac{2| f'(t)|+\epsilon_1 }{f_n(t)}\\
    &\le \sup_{|t| \le k_n} \frac{2| f'(t)|+\epsilon_1 }{f(t)-f(t)+f_n(t) }\\
    &\le \sup_{|t| \le k_n} \frac{2| f'(t)|+\epsilon_1 }{f(t) -\epsilon_0} 
    \label{eq:TheUglyBound}\\
    &= \sup_{|t| \le k_n} \frac{2 \left| \frac{f'(t)}{f(t)}\right| + \frac{\epsilon_1}{ f(t)} }{1 -\frac{\epsilon_0}{f(t)}}\\
    &\le \frac{2 \sup_{|t| \le k_n} \left| \frac{f'(t)}{f(t)}\right| + \epsilon_1 \phi(k_n) }{1 -\epsilon_0 \phi(k_n)}, \label{eq:BoundOntheFirstTermOfTheClip}
\end{align}
where the bound in \eqref{eq:TheUglyBound} follows from the assumptions in \eqref{eq:epsilon_i} and the properties of $\phi$ that imply 
\begin{align}
    \epsilon_0 \phi(k_n) < 1  
    &\Rightarrow \frac{ \epsilon_0 }{f(t)} < 1, \forall |t| \le k_n \\
    &\Rightarrow \epsilon_0 < f(t),  \forall |t| \le k_n;
\end{align}
and the bound in \eqref{eq:BoundOntheFirstTermOfTheClip} follows form the definition of $\phi$ in \eqref{eq:Definition of phi}.  
Now consider the second term in \eqref{eq:ClippedDifferenceBound} 
\begin{align}
    \sup_{|t| \le k_n}  \frac{ 1 }{ f_n(t)}
    & =  \sup_{|t| \le k_n}  \frac{ 1 }{  f_n(t)-f(t) +f(t)}\\
    & \le  \sup_{|t| \le k_n}  \frac{ 1 }{   f(t) -\epsilon_0} \\
    & =  \sup_{|t| \le k_n}  \frac{ 1 }{   1- \frac{\epsilon_0}{f(t)} } \frac{1}{f(t)}
    \label{eq:TheUglyBound2}\\
    &\le \frac{ 1 }{   1- \epsilon_0 \phi(k_n) } \phi(k_n),  \label{eq:FinalClipBOund}
\end{align}
where \eqref{eq:TheUglyBound2} follows by using similar steps leading to the bound in \eqref{eq:TheUglyBound}; and \eqref{eq:FinalClipBOund} follows from the definition of $\phi$.   

Combining the bounds in \eqref{eq:DiffOfFishersManipulation}, \eqref{eq:ClippedDifferenceBound},  \eqref{eq:BoundOntheFirstTermOfTheClip}, and \eqref{eq:FinalClipBOund} concludes the proof.

\section{A Proof of Theorem~\ref{thm:modFIestError}}
\label{app:thm_modFIestError}

Before proceeding with the proof we make several observations. 
First, assumption \eqref{eq:epsilon_phi} implies that
\begin{align}
\inf_{|t| \le k_n} f(t) > \epsilon_0. \label{eq:f>epsilon0}
\end{align}
Second, from the assumptions in \eqref{eq:sup_f} and \eqref{eq:epsilon_i}, one obtains that
\begin{align}
    \sup_{|t| \le k_n} f_n(t) \le f_0 + \epsilon_0. \label{eq:sup_fn}
\end{align}
Third, from the assumptions in \eqref{eq:epsilon_i} and \eqref{eq:f>epsilon0}, we have that
\begin{align}
    \sup_{|t| \le k_n} \frac{1}{f_n(t)} 
    &\le \sup_{|t| \le k_n} \frac{1}{f(t) - \epsilon_0} \\
    &\le \frac{ \phi(k_n)}{ 1- \epsilon_0 \phi(k_n)},
    \label{eq:sup_1/fn}
\end{align}
where the inequality in \eqref{eq:sup_1/fn} follows from the definition of $\phi$.  Finally, by combining \eqref{eq:sup_fn} and \eqref{eq:sup_1/fn} we have that 
\begin{align}
    |\log(f_n)| &\le \max \left(   \log(f_n),   \log \left( \frac{1}{f_n} \right) \right) \\ 
    & \le  \max \left( \log(f_0 + \epsilon_0), \log\left(\frac{\phi(k_n)}{1-\epsilon_0 \phi(k_n)}\right) \right) \\
    & \le \psi(\epsilon_0,k_n). \label{eq:BoundABSlog}
\end{align}

Now,  using the triangle inequality we have that 
\begin{align}
    \left |I(f) - I_{n} \right| 
    &\le \left| \int_{|t| \le k_n} \frac{(f_n'(t))^2}{f_n(t)} - \frac{(f'(t))^2}{f(t)} \dr t  \right| + c(k_n).  
    \label{eq:DiffFIManipulation}
\end{align}
Next, we bound the first term in \eqref{eq:DiffFIManipulation} 
\begin{align}
    &\left| \int_{|t| \le k_n} \frac{(f_n'(t))^2}{f_n(t)}- \frac{(f'(t))^2}{f(t)} \dr t  \right| \notag\\
    &= \left| \int_{|t| \le k_n} \frac{f(t) (f_n'(t))^2- f_n(t) (f'(t))^2}{ f_n(t) f(t)} \dr t \right| \\
    &\leq \left| \int_{|t| \le k_n} \frac{f(t) (f_n'(t))^2- f(t) (f'(t))^2}{ f_n(t) f(t)} \dr t \right| + \left| \int_{|t| \le k_n} \frac{f(t) (f'(t))^2- f_n(t) (f'(t))^2}{f_n(t) f(t)} \dr t \right| \\
    &= \left| \int_{|t| \le k_n} \frac{ (f_n'(t))^2- (f'(t))^2 }{ f_n(t)}  \dr t  \right| + \left| \int_{|t| \le k_n} \frac{ f_n(t)-f(t)}{ f_n(t)} \frac{(f'(t))^2}{f(t)} \dr t \right|\\
    &\le \epsilon_1 \int_{|t| \le k_n} \left| \frac{ f_n'(t) + f'(t) }{ f_n(t)} \right| \dr t + \epsilon_0 \int_{|t| \le k_n} \left| \frac{(f'(t))^2}{ f_n(t) f(t)} \right| \dr t  
    \label{eq:pluginepsilons}\\
    &\le \epsilon_1 \int_{|t| \le k_n} \left| \frac{ f_n'(t) }{ f_n(t)} \right|  \dr t  + \epsilon_1 \int_{|t| \le k_n} \left| \frac{ f'(t) }{ f_n(t)} \right|  \dr t   + \epsilon_0 \rho_{\max}(k_n) \int_{|t| \le k_n} \left| \frac{f'(t)}{ f_n(t)} \right| \dr t  
    , \label{eq:ClippedDiffBound}
\end{align}
where the inequality in \eqref{eq:pluginepsilons} follows from the assumptions in \eqref{eq:epsilon_i}, and the last bound follows from the triangle inequality together with the definition of $\rho_{\max}$.

Now consider the integral in the first term in \eqref{eq:ClippedDiffBound} 
\begin{align}
    \int_{|t| \le k_n} \left| \frac{ f_n'(t) }{ f_n(t)} \right|  \dr t 
    &= \int_{|t| \le k_n} \left| \nabla \log(f_n(t)) \right|  \dr t \\
    &= \int_{|t| \le k_n} \sign\left(\nabla \log(f_n(t)) \right ) \cdot \nabla \log(f_n(t))  \dr t \\
    &= \sign\left(\nabla \log(f_n(t)) \right) \cdot \log(f_n(t)) \Big|_{-k_n}^{k_n} \notag\\
    &\qquad - \int_{|t| \le k_n} \log(f_n(t)) \frac{\dr}{\dr t}\sign\left(\nabla \log(f_n(t)) \right) \dr t, 
    \label{eq:intbypart}
\end{align}
where the inequality in \eqref{eq:intbypart} follows from integration by parts. 
The first term in \eqref{eq:intbypart} can be upper bounded as 
\begin{align}
    &  \sign\left(\nabla \log(f_n(t)) \right) \cdot \log(f_n(t)) \Big|_{-k_n}^{k_n} \
    \le  2 \psi(\epsilon_0,k_n), 
    \label{eq:bound_abs_log}
\end{align}
where the inequality in \eqref{eq:bound_abs_log} follows from \eqref{eq:BoundABSlog}.
In addition, the second term in \eqref{eq:intbypart} is given by  
\begin{align}
    - \int_{|t| \le k_n} \log(f_n(t)) \frac{\dr}{\dr t}\sign\left(\nabla \log(f_n(t)) \right) \dr t 
    &= - \sum_{t \in [-k_n,k_n]: f'_n(t)=0} \log(f_n(t)) \\
    &\le d_{f_n}(k_n) \psi(\epsilon_0,k_n).
    \label{eq:bound_sumlog}
\end{align}
By substituting \eqref{eq:bound_abs_log} and \eqref{eq:bound_sumlog} into \eqref{eq:intbypart}, one obtains
\begin{align}
    \int_{|t| \le k_n} \left| \frac{ f_n'(t) }{ f_n(t)} \right|  \dr t 
    \le (2+d_{f_n}) \psi(\epsilon_0,k_n) .
    \label{eq:bound_int_rhon}
\end{align}

Next, we consider $\int_{|t| \le k_n} \left| \frac{f'(t)}{ f_n(t)} \right| \dr t$ common to the second and the third terms in \eqref{eq:ClippedDiffBound} 
\begin{align}
    \int_{|t| \le k_n} \left| \frac{ f'(t) }{ f_n(t)} \right|  \dr t 
    &\le \int_{|t| \le k_n} \left| \frac{ f'(t) }{ f(t) - \epsilon_0} \right|  \dr t \label{eq:boundf'/fn}\\
    &= \int_{|t| \le k_n} \left| \nabla \log(f(t) - \epsilon_0) \right|  \dr t \\
    &= \int_{|t| \le k_n} \sign\left(\nabla \log(f(t) - \epsilon_0) \right ) \cdot \nabla \log(f(t) - \epsilon_0) \dr t \\
    &= \sign\left(\nabla \log(f(t) - \epsilon_0) \right) \cdot \log(f(t) - \epsilon_0) \Big|_{-k_n}^{k_n} \notag\\
    &\qquad - \int_{|t| \le k_n} \log(f(t) - \epsilon_0) \frac{\dr}{\dr t}\sign\left(\nabla \log(f(t) - \epsilon_0) \right) \dr t \\
    &\le (2+d_f(k_n)) \max \left( \log(f_0 - \epsilon_0), \log\left(\frac{\phi(k_n)}{1-\epsilon_0 \phi(k_n)}\right) \right)\label{eq:bound_int_df/fn} \\
    &\le (2+d_f(k_n)) \psi(\epsilon_0,k_n), 
\end{align}
where the inequalities in \eqref{eq:boundf'/fn} follows from the assumptions in \eqref{eq:epsilon_i} and \eqref{eq:f>epsilon0}, and the bound in \eqref{eq:bound_int_df/fn} follows by using similar step leading to the bound in \eqref{eq:bound_int_rhon}.   

Combining the bounds in \eqref{eq:DiffFIManipulation}, \eqref{eq:ClippedDiffBound},  \eqref{eq:bound_int_rhon} and \eqref{eq:bound_int_df/fn} concludes the proof. 

\section{A Proof of Theorem~\ref{thm:clippedFIestError} } 
\label{app:thm_clippedFIestError}

The difficulty in bounding the error of a clipped estimator is in showing that the clipping is strict enough to avoid gross overestimation, yet permissive enough to avoid gross underestimation. The proof presented here is based on two auxiliary estimators that are constructed to under- and overestimate $I_n^\text{c}(f_n)$ in a controlled manner.
    
Let
\begin{equation}
    \underline{I}_n 
    = \int_{-k_n}^{k_n} \frac{\lceil f'_n(t) - \epsilon_1 \rfloor^2}{f_n(t) + \epsilon_0} \,\dr t,
\end{equation}
where $\lceil \bullet - \epsilon \rfloor$ denotes an ``$\epsilon$-compression'' operator, \textit{i.e.},
\begin{align}
    \lceil f(t) - \epsilon \rfloor 
    = \begin{cases}
        f(t) - \epsilon, & f(t) > \epsilon \\
        0, & - \epsilon \leq f(t) \leq \epsilon \\
        f(t) + \epsilon, & f(t) < -\epsilon.
    \end{cases}
\end{align}
Next, consider the estimator
\begin{equation}
    \overline{I}_n 
    = \int_{-k_n}^{k_n} \frac{\lceil f'_n(t) - \gamma_{1,n}(t) \rfloor^2}{f_n(t) + \gamma_{0,n}(t)} \,\dr t,
\end{equation}
where the functions $\gamma_{i,n} \colon \mathbb{R} \to [0, \epsilon_i]$, $i = 0,1$ are chosen as follows: If it holds that
\begin{equation}
    \lvert \rho_n(t) \rvert \leq \lvert \overline{\rho} (t) \rvert,
\end{equation}
then $\gamma_{0,n}(t) = \gamma_{1,n}(t) = 0$. If, on the other hand,
\begin{equation}
    \lvert \rho_n(t) \rvert > \lvert \overline{\rho} (t) \rvert,
\end{equation}
then $\gamma_{0,n}(t)$ and $\gamma_{1,n}(t)$ are chosen such that
\begin{equation}
    \frac{\lceil f'_n(t) - \gamma_{1,n}(t) \rfloor}{f_n(t) + \gamma_{0,n}(t)} = \overline{\rho} (t).
\end{equation}
Note that since
\begin{equation}
    \left\lvert \frac{\lceil f'_n(t) - \epsilon_1 \rfloor}{f_n(t) + \epsilon_0} \right\rvert \leq \lvert \rho(t) \rvert \leq \lvert \overline{\rho} (t) \rvert,
\end{equation}
this is always possible.
    
In Appendix~\ref{app:ProofofLemmaRegFIerror} it is shown that the following relations hold between the estimators defined above:
\begin{align}
    \underline{I}_n &\leq I(f), 
    \label{eq:lower_bound_I} \\
    \underline{I}_n &\leq I_n^\text{c}, 
    \label{eq:lower_bound_Iclipped} \\
    I_n^\text{c} &\leq \overline{I}_n + \epsilon_1 \Phi_\text{max}^1(k_n),
    \label{eq:diff_Iclipped_Ilow} \\
    I(f) - \underline{I}_n &\leq 4 \epsilon_1 \Phi^1(k_n) + 2 \epsilon_0 \Phi^2(k_n) + c(k_n), 
    \label{eq:diff_I_Ilow} \\
    \overline{I}_n - \underline{I}_n &\leq 2 \epsilon_1 \Phi_\text{max}^1(k_n) + \epsilon_0 \Phi_\text{max}^2(k_n). \label{eq:diff_Ihi_Ilow}
\end{align}

The bound in Theorem~\ref{thm:clippedFIestError} can now be obtained by bounding the under- and overestimation errors separately. For $I_n^\text{c} \leq I(f)$ it holds that
\begin{align}
    I(f) - I_n^\text{c} &\leq I(f) - \underline{I}_n \\
    &\leq 4 \epsilon_1 \Phi^1(k_n) + 2 \epsilon_0 \Phi^2(k_n)  + c(k_n).
\end{align}
For $I_n^\text{c} > I(f)$ it hold that 
\begin{align}
    I_n^\text{c} - I(f) &\leq \overline{I}_n - \underline{I}_n + \epsilon_1 \Phi_\text{max}^1(k_n) \\
    &\leq 3 \epsilon_1 \Phi_\text{max}^1(k_n) +  \epsilon_0 \Phi_\text{max}^2(k_n).
\end{align}   
The bound in \eqref{eq:diffFI_clipped} follows. Furthermore, following the same steps as those leading to the bound in \eqref{eq:bound_int_rhon}, the bound in \eqref{eq:diffFI_clipped_mod} follows.

\section{A Proof of Estimator Relations in Theorem~\ref{thm:clippedFIestError} }
\label{app:ProofofLemmaRegFIerror}

The bound in \eqref{eq:lower_bound_I} follows directly from the fact that under the assumptions in \eqref{eq:epsilon_i} 
\begin{equation}
    \frac{\lceil f'_n(t) - \epsilon_1 \rfloor^2}{f_n(t) + \epsilon_0} \leq \frac{(f'(t))^2}{f(t)}.
\end{equation}
Analogously, \eqref{eq:lower_bound_Iclipped} follows from 
\begin{equation}
    \frac{\lceil f'_n(t) - \epsilon_1 \rfloor^2}{f_n(t) + \epsilon_0} \leq \lvert \rho(t) \rvert \lvert \lceil f'_n(t) - \epsilon_1 \rfloor \rvert \leq \lvert \rho(t) \rvert \lvert f'_n(t) \rvert
\end{equation}
In order to show \eqref{eq:diff_I_Ilow}, note that under the assumptions in \eqref{eq:epsilon_i} it holds that
\begin{align}
  f_n(t) + \epsilon_0 &\geq f(t), \\
  \lvert \lceil f'_n(t) - \epsilon_1 \rfloor \rvert &\leq \lvert f'(t) \rvert, \\
  (f_n(t) + \epsilon_0) - f(t) &\leq 2\epsilon_0, \label{eq:diff0} \\
  \lvert \lceil f'_n(t) - \epsilon_1 \rfloor - f'(t) \rvert &\leq 2\epsilon_1. \label{eq:diff1}
\end{align}
Hence, in analogy to Theorem~\ref{thm:FIestError}, the estimation error of $\underline{I}_n$ can be written as
\begin{equation}
  I(f) - \underline{I}_n = \int_{-k_n}^{k_n} \frac{(f'(t))^2}{f(t)} - \frac{\lceil f'_n(t) - \epsilon_1 \rfloor^2}{f_n(t) + \epsilon_0} \,\dr t + c(k_n).
  \label{eq:est_error_regularized}
\end{equation}
Using the same arguments as in the proof of Theorem~\ref{thm:FIestError}, the integral term on the right hand side of \eqref{eq:est_error_regularized} can be bounded by
\begin{align}
  &\! \int_{-k_n}^{k_n} \frac{(f'(t))^2}{f(t)} - \frac{\lceil f'_n(t) - \epsilon_1 \rfloor^2}{f_n(t) + \epsilon_0} \,\dr t \notag\\ 
  &= \left\lvert \int_{-k_n}^{k_n} \frac{\lceil f'_n(t) - \epsilon_1 \rfloor^2 f(t) - (f'(t))^2 (f_n(t) + \epsilon_0)}{f(t)(f_n(t) + \epsilon_0)} \,\dr t \right\rvert  \\
  &= \left\lvert \int_{-k_n}^{k_n} \frac{\lceil f'_n(t) - \epsilon_1 \rfloor^2 f(t) - (f'(t))^2 (f_n(t) + \epsilon_0)}{f(t)(f_n(t) + \epsilon_0)} \,\dr t \right\rvert  \\
  &\leq \left\lvert \int_{-k_n}^{k_n} \left\lvert \lceil f'_n(t) - \epsilon_1 \rfloor - f'(t)\right\rvert  \frac{\left\lvert \lceil f'_n(t) - \epsilon_1 \rfloor + f'(t) \right\rvert}{f_n(t) + \epsilon_0} \,\dr t \right\rvert \notag \\ 
  &\qquad + \int_{-k_n}^{k_n} \left\lvert f(t) - (f_n(t) + \varepsilon_0) \right\rvert  \frac{(f'(t))^2}{f(t)(f_n(t) + \varepsilon_0)} \,\dr t \\
  &\leq 2 \epsilon_1 \int_{-k_n}^{k_n} \frac{\left\lvert \lceil f'_n(t) - \epsilon_1 \rfloor \right\rvert + \lvert f'(t) \rvert}{f_n(t) + \epsilon_0} \,\dr t  + 2 \epsilon_0 \int_{-k_n}^{k_n} \frac{(f'(t))^2}{f(t)(f_n(t) + \epsilon_0)} \,\dr t \\
  &\leq 2 \epsilon_1 \int_{-k_n}^{k_n} 2 \left\lvert \frac{f'(t)}{f(t)} \right\rvert \,\dr t + 2 \epsilon_0 \int_{-k_n}^{k_n} \left\lvert \frac{f'(t)}{f(t)} \right\rvert^2 \,\dr t \\
  &\leq 4 \epsilon_1 \int_{-k_n}^{k_n} \left\lvert \rho(t) \right\rvert + 2 \epsilon_0 \int_{-k_n}^{k_n} \rho^2(t) \,\dr t \\
  &= 4 \epsilon_1 \; \Phi^1(k_n) + 2 \varepsilon_0 \; \Phi^2(k_n).
\end{align}
Using the same steps, it is not difficult to show \eqref{eq:diff_Ihi_Ilow}, 
where the factor $2$ does not arise since, in contrast to \eqref{eq:diff0} and \eqref{eq:diff1},
\begin{align}
    \lceil f_n(t) + \epsilon_0 \rfloor - \lceil f_n(t) + \gamma_{0,n}(t) \rfloor &\leq \epsilon_0, \\
    \lceil f'_n(t) - \gamma_{1,n}(t) \rfloor - \left\lceil f'_n(t) - \epsilon_1 \right\rfloor &\leq \epsilon_1,
\end{align}
and $c(k_n)$ does not arise since both estimators are defined on $[-k_n, k_n]$. 

In order to show \eqref{eq:diff_Iclipped_Ilow}, first note that for $\lvert \rho_n(t) \rvert \leq \lvert \overline{\rho}(t) \rvert$ it holds that
\begin{equation}
    \frac{\lceil f'_n(t) - \gamma_{1,n}(t)\rfloor ^2}{f_n(t)+ \gamma_{0,n}(t)} 
    = \frac{(f'_n(t))^2}{f_n(t)} 
    = \lvert \rho_n(t) \rvert \lvert f'_n(t) \rvert,
\end{equation}
\textit{i.e.}, $\overline{I}_n(f_n) = I^\text{c}_n(f_n)$. 
Hence, $I_n^\text{c}(f_n) > \overline{I}(f_n)$ implies $\lvert \rho_n(t) \lvert \geq \lvert \overline{\rho}(t) \lvert$ on some region of $[-k_n, k_n]$. On this region it holds that
\begin{align}
    &\!\frac{\lceil f'_n(t) - \gamma_{1,n}(t) \rfloor^2}{f_n(t) + \gamma_{0,n}(t)} \notag \\
    &= \frac{\lvert \lceil f'_n(t) - \gamma_{1,n}(t) \rfloor \rvert}{f_n(t) + \gamma_{0,n}(t)} \lvert \lceil f'_n(t) - \gamma_{1,n}(t) \rfloor \rvert \\
    &= \lvert \overline{\rho}(t) \rvert \, \lvert \lceil f'_n(t) - \gamma_{1,n}(t) \rfloor \rvert.
\end{align}
Since
\begin{equation}
    \lvert f'_n(t) \rvert - \lvert \lceil f'_n(t) - \gamma_{1,n}(t) \rfloor \rvert \leq \gamma_{1,n} \leq \epsilon_1
\end{equation}
it follows that
\begin{align}
    I_n^\text{c}(f_n) - \overline{I}_n(f_n) &\leq \int_{-k_n}^{k_n} \lvert \overline{\rho}(t) \rvert \epsilon_1 \,\dr t  \\
    &\leq \epsilon_1 \Phi_\text{max}^1(k_n).
\end{align}

\section{A Proof of Lemma~\ref{lem:FIestErrGaussianNoise}} 
\label{prof:lem:FIestErrGaussianNoise}

We begin by bounding $v_r$ and $\delta_{r,a}$. First, 
\begin{align}
    v_0&= \int | t | k(t) \dr t= \sqrt{\frac{2}{\pi}},\\
    v_1&= \int \left| t^2-1 \right| k(t) \dr t= 2 \sqrt{\frac{2}{ {\rm e}\pi}}.
\end{align}
Second, 
\begin{align}
    \delta_{r,a}
    &= \left| \mathbb{E}[f_n^{(r)}(t)]-f_Y^{(r)}(t) \right | \\
    &=  \left| \int  \frac{1}{a}   k \left( \frac{t-y}{a} \right)  \left(  f_Y^{(r)}(y) -f_Y^{(r)}(t) \right)\dr y  \right| \\
    &=   \left| \int     k \left( y \right)  \left(  f_Y^{(r)}(t+a y) -f_Y^{(r)}(t) \right)\dr y   \right| \\
    & \le   \sup_{t \in \mathbb{R}} \left| f_Y^{(r+1)}(t)  \right| \int     k \left( y \right)  a  |y|  \dr y      \\
    &= a \sqrt{ \frac{2}{\pi}}     \sup_{t \in \mathbb{R}} \left| f_Y^{(r+1)}(t)  \right| . 
    \label{eq:BoundOnDifferenceOfExpectedValue}
 \end{align} 
Now, for $r=0$,
\begin{align}
    \left| f_Y^{(1)}(t)  \right| 
    &= \left| \mathbb{E} \left[ (t- \sqrt{\snr}X) \frac{1}{\sqrt{2 \pi }} {\rm e}^{- \frac{(t-\sqrt{\snr}X)^2}{2}}  \right] \right|\\
    & \le \frac{1}{\sqrt{2 \pi }} \frac{1}{\sqrt{\rm e}}, 
    \label{eq:BoundOnFirstDerivativeOfOutpdf}
\end{align}
where we have used the bound $t {\rm e}^{-\frac{t^2}{2}} \le \frac{1}{\sqrt{\rm e}}$.  For $r=1$,
\begin{align}
    \left| f_Y^{(2)}(t) \right| 
    &= \left| \mathbb{E} \left[ \left( (t-\sqrt{\snr}X)^2-1 \right) \frac{1}{\sqrt{2 \pi }} {\rm e}^{- \frac{(t- \sqrt{\snr}X)^2}{2}}  \right]  \right|\\
    & \le \frac{1}{\sqrt{2 \pi }} \frac{2}{ {\rm e}}+\frac{1}{\sqrt{2 \pi }}, 
    \label{eq:BoundOnSecondDerivativeOfOutpdf}
\end{align}
where we have used the bound $t^2 {\rm e}^{-\frac{t^2}{2}}  \le \frac{2}{\rm e}$. 

Next, we bound the score function $\rho_Y$
\begin{align}
    |\rho_Y(t)|
    &= \left| \frac{f_Y'(t)}{f_Y(t)} \right| \\
    &= \left| \sqrt{\snr} \E[X|Y=t] - t \right| 
    \label{eq:scoreFunctionandConditionalExpectation} \\
    &\le \sqrt{\snr} \E \left[  \left| X \right||Y=t \right] + |t|\\
    &\le \sqrt{\snr} \sqrt{\E \left[  X^2|Y=t \right] } + |t| 
    \label{eq:JensensInequalityToConditional}\\
    & \le \sqrt{3 \snr \var(X) +  4 t^2 } + |t| 
    \label{eq:BoundOnCOnditinalExpectation}\\
    & \le \sqrt{3 \snr \var(X) } + 3|t|, 
    \label{eq:BoundOnTheScoreFunction}
\end{align} 
where the equality in \eqref{eq:scoreFunctionandConditionalExpectation} follows by using the identify $\frac{f_Y'(t)}{f_Y(t)}=\sqrt{\snr} \E[X|Y=t] -t$ \cite{esposito1968relation}; the inequality in \eqref{eq:JensensInequalityToConditional} follows from Jensen's inequality; and the inequality in \eqref{eq:BoundOnCOnditinalExpectation} follows from the bound in \cite[Proposition 1.2]{fozunbal2010regret}.  
Using the bound in \eqref{eq:BoundOnTheScoreFunction} it follows that 
\begin{align}
    \rho_{\max}(k_n)
    &=\max_{|t|\le k_n} |\rho(t)|  \le \sqrt{3 \snr \var(X) } +3 k_n. 
\end{align}
Using the relation between the Fisher information and the MMSE, we have that
\begin{align}
    I(f_Y)& = 1-\snr \mmse(X,\snr) \le 1. \label{eq:FImmse}
\end{align}
Finally, the function $\phi$ is obtained by observing that
\begin{align}
    f_Y(t)&= \E \left[ \frac{1}{\sqrt{2 \pi}} {\rm e}^{-\frac{(t-\sqrt{\snr}X)^2}{2}} \right]\\
    & \ge \frac{1}{\sqrt{2 \pi}} {\rm e}^{-\frac {\E \left[(t-\sqrt{\snr}X)^2 \right]}{2}} \\
    &\ge  \frac{1}{\sqrt{2 \pi}} {\rm e}^{- \left(t^2 + \snr \E[X^2]\right)} ,
\end{align}
where we used Jensen's inequality and the fact that $(a+b)^2 \le 2 (a^2+b^2)$.

This concludes the proof. 

\section{A Proof of Lemma~\ref{lem:ckn}} 
\label{app:lem_ckn}

Choose some $v>0$. Then
\begin{align}
    c(k_n)
    &= \E \left[ \rho_Y^2(Y)  1_{\{|Y| \ge k_n \}} \right] \\
    &\le \E^{\frac{1}{1+v}} \left[ |\rho_Y(Y)|^{2(1+v)} \right] \mathbb{P}^{\frac{v}{1+v}} \left[|Y| \ge k_n \right] 
    \label{eq:HoldersInequality} \\
    &= \E^{\frac{1}{1+v}} \left[ |\E[Z|Y]|^{2(1+v)} \right] \mathbb{P}^{\frac{v}{1+v}} \left[|Y| \ge k_n \right] 
    \label{eq:IdentityForScoreFunction} \\
    &\le \E^{\frac{1}{1+v}} \left[ |Z|^{2(1+v)} \right] \mathbb{P}^{\frac{v}{1+v}} \left[|Y| \ge k_n \right] \\
    &= \frac{2 \Gamma^{ \frac{1}{(1+v)}} \left(v+\frac{1}{2} \right)}{ \pi^{ \frac{1}{2(1+v)}}} \mathbb{P}^{\frac{v}{1+v}} \left[|Y| \ge k_n \right] \\
    &= \frac{2 \Gamma^{ \frac{1}{(1+v)}} \left(v+\frac{1}{2} \right)}{ \pi^{ \frac{1}{2(1+v)}}} \left(  \frac{ \snr \E[|X|^2] +1}{k_n^2} \right)^{\frac{v}{1+v}}, 
    \label{eq:MarkovInequality}
\end{align}
where \eqref{eq:HoldersInequality} follows from H\"older's inequality; \eqref{eq:IdentityForScoreFunction} follows by using the identity 
\begin{align}
    \rho_Y(t)=\sqrt{\snr} \E[X|Y=t] -t= -\E[Z|Y=t ];
\end{align}
and \eqref{eq:MarkovInequality} follows from Markov's inequality. 
 
Now, if $\E[X^2]<\infty$, then using Markov's inequality
\begin{align}
    \mathbb{P} \left[|Y| \ge k_n \right] 
    &\le \frac{\E[Y^2]}{k_n^2}
    =\frac{\snr \E[|X|^2] +1}{k_n^2}.
\end{align}

Moreover, using the Chernoff bound,
\begin{align}
    \mathbb{P} \left[|Y| \ge k_n \right] 
    & \le  {\rm e}^{-k_n t}  \E \left[{\rm e}^{t |Y|}\right]\\
    &\le 2 {\rm e}^{-k_n t +\frac{t^2}{2}}  \E \left[{\rm e}^{t \sqrt{\snr}|X|}\right]\\
    &= 2 {\rm e}^{-k_n t +\frac{t^2}{2}}  {\rm e}^{ \frac{\alpha^2 \snr}{2}}.
\end{align}
Therefore,
\begin{align}
    c (k_n)  
    &\le \inf_{t>0}  \inf_{ v >0} \frac{2 \Gamma^{ \frac{1}{(1+v)}} \left(v+\frac{1}{2} \right)}{ \pi^{ \frac{1}{2(1+v)}}}  2^{\frac{v}{1+v} } {\rm e}^{ \frac{v}{1+v}  \left(-k_n t +\frac{t^2}{2}+\frac{\alpha^2 \snr}{2} \right)}  \\
    &\le \inf_{ v >0} \frac{2 \Gamma^{ \frac{1}{(1+v)}} \left(v+\frac{1}{2} \right)}{ \pi^{ \frac{1}{2(1+v)}}} 2^{ \frac{v}{1+v} } {\rm e}^{\frac{v}{1+v} \frac{\alpha^2 \snr - k_n^2}{2} }. 
\end{align}
This concludes the proof.

\section{A Proof of Theorem~\ref{thm:FIestError_GaussianNoise}}
\label{app:thm_FIerr_GaussianNoise}

Let
\begin{align}
    \varepsilon_n &= \frac{4  \epsilon k_n \rho_{\max}(k_n) +  \epsilon\phi(k_n) + 2 \epsilon^2  k_n \phi(k_n) }{1 -\epsilon \phi(k_n)} + c(k_n).
\end{align}
To apply the bounds in Theorem~\ref{thm:boundsEstDensityandDerivatives} and Theorem~\ref{thm:FIestError}, the following equalities/inequalities must hold for $r \in \{ 0,1 \}$:
\begin{subequations}\label{eq:assumptions}
\begin{align}
    \epsilon &> \delta_{r,a}, \\
    \frac{a^{2r+2} (\epsilon_r - \delta_{r,a})^2}{v_r^2} &\gg \frac{1}{n}, \\
    \lim_{n\to\infty} \epsilon^2  k_n \phi(k_n) &= 0, \\
    \lim_{n\to\infty} \epsilon \phi(k_n) &= 0,\\
    \lim_{n\to\infty} c(k_n) &= 0.
\end{align}
\end{subequations}
To satisfy \eqref{eq:assumptions}, we choose
\begin{align}
    a &= n^{-w}, w \in \left(0,\frac{1}{6} \right), \label{eq:a} \\
    k_n &= \sqrt{ u \log(n) }, u \in \left(0,w \right), \label{eq:kn}\\
    \epsilon &= a. \label{eq:epsilon}
\end{align}
Then, together with the bounds in Lemma~\ref{lem:FIestErrGaussianNoise}, the relevant quantities in \eqref{eq:assumptions} are as follows:
\begin{subequations}\label{eq:pluginkna}
\begin{align}
    \frac{a^2 (\epsilon_0 - \delta_{0,a})^2}{v_0^2} &= c_1 n^{- 4 w}, \\
    \frac{a^4 (\epsilon_1 - \delta_{1,a})^2}{v_1^2} &= c_2 n^{- 6 w}, \\
    \epsilon^2  k_n \phi(k_n) &\le c_5 n^{u-2w} \sqrt{u \log (n)}, \\
    \epsilon \phi(k_n) &\le c_5 n^{u-w},\\
   c(k_n) &\le \frac{c_4}{\sqrt{ u \log(n) }} ,
\end{align}
\end{subequations}
which yields \eqref{eq:boundvarepsilon}. Now, if $|X|$ is $\alpha$-sub-Gaussian, the bound in \eqref{eq:FIerrBound_subGaussian} can be obtained from Lemma~\ref{lem:ckn} with $v=1$. 

Since \eqref{eq:epsilon_i} leads to \eqref{eq:boundDiffFI}, one obtains
\begin{align}
    &\mathbb{P} \left[ \left| I_n(f_n) - I(f_Y) \right| \ge \varepsilon_n \right] \notag\\
    &\le \mathbb{P} \left[ \sup_{ |t| \le k_n} \left| f_n(t) - f_Y(t) \right| \ge \epsilon \right]  + \mathbb{P} \left[ \sup_{ |t| \le k_n} \left| f'_n(t) - f'_Y(t) \right| \ge \epsilon \right] \\
    &\le \mathbb{P} \left[ \sup_{t \in \mathbb{R}} \left| f_n(t)-f_Y(t) \right| >\epsilon  \right] + \mathbb{P} \left[ \sup_{t \in \mathbb{R}} \left| f'_n(t)-f'_Y(t) \right| >\epsilon  \right]\\
    &\le 2 {\rm e}^{- n \pi a^{2} \left(\epsilon- a \frac{1}{\sqrt{2 \pi {\rm e}}}  \right)^2 } + 2 {\rm e}^{- n {\rm e} \pi a^{4} \left(\epsilon- a \frac{\frac{2}{{\rm e}}+1}{\sqrt{2 \pi }} \right)^2 } 
    \label{eq:applyThEstDerivative}\\
    &= 2 {\rm e}^{-  \pi \left(1 - \frac{1}{\sqrt{2 \pi {\rm e}}}  \right)^2   n^{1-4w} } + 2 {\rm e}^{- {\rm e} \pi \left(1 - \frac{\frac{2}{{\rm e}}+1}{\sqrt{2 \pi}} \right)^2 n^{1-6w} }, 
    \label{eq:bound_prob_err}
\end{align}
where the inequality in \eqref{eq:applyThEstDerivative} follows from Theorem~\ref{thm:boundsEstDensityandDerivatives}, and the last step follows from \eqref{eq:kn}, \eqref{eq:a}, and \eqref{eq:epsilon}. This concludes the proof.

\section{A Proof of Theorem~\ref{thm:clipped_FIest_GaussianNoise}}
\label{app:thm_clippedFIest}

Let
\begin{align}
    \varepsilon_n 
    &= 8 \epsilon \Phi_1(k_n)  + 4 \epsilon \Phi_2(k_n) + c(k_n).
\end{align}
To apply the bounds in Theorem~\ref{thm:clippedFIestError} and Lemma~\ref{lem:FIestErrGaussianNoise}, the following equalities/inequalities must hold for $r \in \{ 0,1 \}$:
\begin{subequations}\label{eq:assumption}
\begin{align}
    \epsilon &> \delta_{r,a}, \\
    a^{2r+2} (\epsilon - \delta_{r,a})^2 &\gg \frac{1}{n}, \\
    \lim_{n\to\infty} \epsilon \Phi_1(k_n) &= 0, \\
    \lim_{n\to\infty} \epsilon \Phi_2(k_n) &= 0, \\
    \lim_{n\to\infty} c(k_n) &= 0.
\end{align}
\end{subequations}
To satisfy \eqref{eq:assumption}, we choose
\begin{align}
    a &= n^{-w}, w \in \left(0,\frac{1}{4} \right), \label{eq:a_} \\
    k_n &= n^u, u \in \left(0, \frac{w}{3} \right), \label{eq:k_n}\\
    \epsilon &= a. 
\end{align}
Then, together with the bounds in Lemma~\ref{lem:FIestErrGaussianNoise}, the relevant quantities in \eqref{eq:assumption} are as follows:
\begin{subequations}
\begin{align}
    a_r^{2r+2} (\epsilon_r - \delta_{r,a})^2 &= \beta_r n^{ (2r+r) w_r}, r = 0,1, \\
    \epsilon \Phi_1(k_n) &\le 2 n^{u-w} \left( c_3 + 3 n^u \right), \\
    \epsilon \Phi_2(k_n) &\le 2 n^{u-w} \left(c_3 + 3 n^u \right)^2,\\
   c(k_n) &\le c_4 n^{-u} ,
\end{align}
\end{subequations}
which yields \eqref{eq:boundnewFIerr}. Moreover, if $|X|$ is $\alpha$-sub-Gaussian, the bound in \eqref{eq:clippedFIerrBound_subGaussian} can be obtained from Lemma~\ref{lem:ckn}. 

By using steps similar to those leading to \eqref{eq:bound_prob_err}, we have that
\begin{align}
    \mathbb{P} \left[ \left| I_n^\text{c}(f_n) - I(f_Y) \right| \ge \varepsilon_n \right] 
    &\le 2 {\rm e}^{- \pi \left(1 - \frac{1}{\sqrt{2 \pi {\rm e}}} \right)^2 n^{1-4 w} } + 2 {\rm e}^{-{\rm e} \pi \left(1 - \frac{\frac{2}{{\rm e}}+1}{\sqrt{2 \pi }} \right)^2  n^{1-6 w} }.
\end{align}
This concludes the proof.

\end{appendices}

\bibliographystyle{IEEEtran}
\bibliography{ref}

\end{document}